\documentclass[10pt,journal,compsoc]{IEEEtran}
\usepackage[pass]{geometry}
\usepackage{amsmath,amssymb,amsfonts}
\usepackage{algorithmic}
\usepackage{graphicx}
\usepackage{textcomp}
\usepackage{xcolor}
\usepackage{fancyhdr}
\usepackage{hyperref}
\usepackage{color}
\usepackage{soul}
\usepackage{epsfig}

\ifCLASSOPTIONcompsoc
  \usepackage[nocompress]{cite}
\else
  \usepackage{cite}
\fi

\usepackage{array}

\ifCLASSOPTIONcompsoc
  \usepackage[caption=false,font=footnotesize,labelfont=sf,textfont=sf]{subfig}
\else
  \usepackage[caption=false,font=footnotesize]{subfig}
\fi

\usepackage{fixltx2e}

\usepackage{stfloats}

\renewcommand{\paragraph}[1]{\par\noindent\textbf{#1.}}

\begin{document}

\title{Monarch: A Durable Polymorphic Memory For Data Intensive Applications}

\author{Ananth Krishna Prasad,
        Mahdi Nazm Bojnordi}

\IEEEtitleabstractindextext{
\begin{abstract}
3D die stacking has often been proposed to build large-scale DRAM-based caches.
Unfortunately, the power and performance overheads of DRAM limit the efficiency of high-bandwidth memories.
Also, DRAM is facing serious scalability challenges that make alternative technologies more appealing.
This paper examines Monarch, a resistive 3D stacked memory based on a novel reconfigurable crosspoint array called XAM.
The XAM array is capable of switching between random access and content-addressable modes, which enables Monarch (i) to better utilize the in-package bandwidth and (ii) to satisfy both the random access memory and associative search requirements of various applications.
Moreover, the Monarch controller ensures a given target lifetime for the resistive stack.
Our simulation results on a set of parallel memory-intensive applications indicate that Monarch outperforms an {ideal DRAM caching} by $1.21\times$ on average.
For in-memory hash table and string matching workloads, Monarch improves performance up to $12\times$ over the conventional high bandwidth memories.
\end{abstract}

\begin{IEEEkeywords}
Reconfigurable Memory, Data-Intensive, Content Addressable Memory, Data Parallel, Novel Memory Systems
\end{IEEEkeywords}}

\maketitle

\IEEEdisplaynontitleabstractindextext

\IEEEpeerreviewmaketitle

\section{Introduction}
\label{section:intro}
With data being generated at such unfathomable pace and quantity, large memories and cache architectures have become the need for big data processing.
While on-chip memories are fast, they are limited in size due to power and area consumption.
On the other hand, off-chip memories, albeit being large, suffer from low bandwidth and costly off-chip data movement. %~\cite{dally.interconnection}.
Off-chip memories also suffer from the so called \textit{bandwidth wall} due to the pin count limitations %\cite{itrs.manual}.
%
%Conventional SRAM caches incur a lot of area overhead and power consumption, imposing a limitation on their size. The size of SRAM caches cannot exceed beyond few tens of megabytes (SRAM LLC in Ivytown processor is 37.5MB \cite{rusu20145}). Moreover, large caches require expensive colling mechanisms which enhance their cost excessively. Even large scale memory systems cannot provide enough bandwidth. For example, some of the large memory systems with 2TB capacity access less than 10\% of thier capacity in billions of processor cycles \cite{lowe2016use}
%
Instead, 
%The bandwidth problem was alleviated through the introduction of 
the through-silicon via (TSV) technology has enabled 3D die-stacked memories that can offer a high bandwidth in excess of terabytes per second~\cite{amdhbm}. %,mcdram,thermal}.
The primary target of such high bandwidth memory has been in-package DRAM for caching and near-data processing.
% has been the primary target for such high bandwidth memories in designing
% have primarily targeted the DRAM technology for storing data in the 3D layers.
%DRAM has been the primary target of such 
%for  , which  by employing through-silicon vias (TSVs).
%These interface technologies, such as high bandwidth memory (HBM), hybrid memory cube (HMC), and wide-IO (WIO), were mainly targeted towards DRAM \cite{hmc, jedechbm}. 
%This allowed for design of gigascale in package DRAM. These 
%The 3D stacked memories were predominantly used as in-package L4 cache, till the recent advent of Intel's Knights Landing~\cite{mcdram}, which enables reconfiguration of in-package DRAM as either L4 cache or an addressable memory independent of the main memory.

In-package DRAM faces significant challenges for large-scale data-intensive computing.
%However, DRAM offers low scalability and density. 
The latency of accesses to 3D die-stacked DRAM is almost identical to that of the conventional off-chip DRAM.
Therefore, DRAM-based in-package caching becomes complex, costly, and often inefficient~\cite{qureshifundamental,lohefficiently}. %7284066,chop,jevdjic2014unison,young2017dice,jang2016efficient}.
Furthermore, an in-package DRAM consumes significant power due to periodic refreshing, large row activation, and multi-level row buffering throughout the memory chips.
In addition to consuming a significant portion of the system energy, the DRAM power may lead to serious thermal issues in the 3D memories.
In particular, this low energy-efficiency results in concerns for the future of DRAM as the need for low-power high-efficiency data processing is never ending.
Moreover, the performance gap between main memory and conventional data storage has been increasing, which is leading to the rise of in-memory database (iMDB) storage for minimizing the query response time~\cite{xu2015overcoming}.
%Memcached \cite{memcached} and VoltDB \cite{voltdb} are popular and widely used examples of iMDBs.

Prior work such as Jenga~\cite{jenga} have looked at defining application-specific cache hierarchies than rigid ones, which eliminates inefficiencies caused by conflicting needs of different applications. Recent work on profiling google consumer workload~\cite{boroumand2018google} shows that data movement contributes to a significant portion of total energy consumption. The majority of this data movement comes from simple primitives across multiple applications, which are good candidates for processing-in-memory.
This provides good motivation for developing a low-cost reconfigurable memory that allows the software to allocate its resources for caching or in-memory processing.

Researchers have been actively looking into alternative technologies for main memory.
Non-volatile memories like resistive RAM (RRAM) and phase-change memory (PCM) exhibit promising results for the future memory systems. %, such as being persistent, where the data survives power outages.
%Intel has made commercially available its Optane DC persistent memory module, which provides access times on the order of DRAM, while providing persistence.
There have been numerous recent proposals examining 3D stacks with resistive memories. %,mramcmp,3Drram}.
In particular, RRAM has shown extra capabilities for in-situ operations beyond caching and storage, ranging from combinatorial optimization to machine learning and DNA sequence alignment~ \cite{bojnordi2016memristive, prasad2021memristive}.
This makes it interesting to explore enabling in-situ application specific computing as part of the 3D die-stacked memory. %, on top of the existing advantages with using RRAM for caching.
%However, building an RRAM-based in-package memory is a significant challenge.
%In this paper, we identify such challenges in building an RRAM-based in-package memory and propose a viable solution for a reconfigurable RRAM-based in-package memory architecture.
%We propose a novel array structure, called XAM, enabling low-cost CAM/RAM reconfigurability.
%We identify the lifetime issues that might potentially arise with write-intensive applications and propose low-cost techniques that help alleviate and manage such issues within the proposed XAM arrays.
%This enables energy efficient and fast in-package caching while allowing user control over the defined CAM/RAM regions. The XAM array structure also allows for low-cost large-scale associative searching across CAM regions. 

The main contributions of the paper are as follows.
(1) Monarch provides the semantics of interfacing commands for a high-bandwidth resistive memory. We provide the primitives for exploiting cache and search capabilities in high bandwidth resistive memories.
(2) The paper provides a comprehensive design for a novel array structure using 2R cells that supports efficient reconfiguration between CAM and RAM. We also examine novel architectural mechanisms for column and row writes in XAM arrays through diagonal organization.
(3) A novel architectural mechanism for toggle-based vault control is proposed that vastly minimizes the controller complexity while ensuring efficient and highly parallel data movement.
(4) This paper identifies the issues with write endurance and proposes techniques for ensuring a lower bound of lifetime.
%(5) We provide a comprehensive analysis of the proposed architecture for three important classes of data intensive applications while keeping the physical TSV organization the same as that of an in-package DRAM.

\section{Motivation and Challenges}
\label{section:MotivandChallenges}

%\subsection{Motivation}
%\label{section:motivation}

%\textcolor{red}{
Classically, computer systems employ two or three levels of SRAM caches to extract maximal latency and bandwidth benefits of data access.
SRAM caches are usually multi-way set associative with addressable sets containing multiple cache-lines, each one distinguishable by a tag.
%But the resolution of the matching way in such SRAM caches employ parallel 
%Multiple tag comparisons across all the ways of each set is necessary to resolve a match.
%This requires logical complexity owing to the requirement of a comparator per way. Nevertheless, such 
The complexity of tag management may be afforded for the SRAM caches due to their low set-associativity and capacity.
% of SRAM caches.
%
%\textcolor{red}{The recent advent of in-package 3D stacked DRAM memory has helped move data closer to memory, with such 3D memories capabile of bandwidth magnitudes higher than traditional off-chip DRAM. Unfortunately, even several gigabytes of such 3D die-stacked DRAM is insufficient to satisfy present age big data workloads' capacity requirements. This has led researchers to explore the alternative of using 3D die-stacked DRAM as large last-level caches . However, such 
However, in 3D die-stacked DRAM caches, tag management remains a significant challenge.
% for 
%
%The recently proposed DRAM caches, though they offer good performance, don't address the source of fundamental challenges such as Tag access overhead, poor hit latency and hit ratios. These issues  can be traced back to DRAM caches being unable to support large multi-way set associative caches. 
Although there has been research on alleviating this issue~\cite{7284066, Accord} DRAM caches suffer the absence of a structure that supports high set-associativity at low latencies.
%, 5375363, jevdjic2013stacked}

%\textcolor{red}{Content addressable memories, or CAMs, offer a natural solution for such large set-associative caches to operate at low tag access overheads and hit latencies at low powers. However, it is a major challenge to incorporate such CAM structures in conventional 3D die-stacked DRAM memories}

%\textcolor{red}{
Content addressable memory (CAM) offers a natural solution for tag comparison in large set-associative caches.
However, incorporating CAM in the conventional 3D die-stacked DRAM is a major challenge.
Unlike DRAM, RRAM has been shown capable of large-scale in-situ parallel bitwise XOR/XNOR \cite{rcache}, which makes it an ideal choice for large-scale CAM designs.
%, ni2017energy
%Such a RRAM based 3D-stacked memory could potentially majorly alleviate the high tag overheads, high hit latency and low hit ratios associated with DRAM caches.}
%
%\textcolor{red}{
Prior work has made such CAM structure visible to the cache controller only~\cite{rcache}.
%, which ensures proper tagging of cacheblocks and tag lookups in such CAM structures. 
Interestingly, the inherent nature of CAM structures make them suitable for indexing applications requiring large-scale parallel searching, such as hash tables and string matching.
Large scale parallel index-based searching is a very prevalent application in database systems~\cite{manolopoulos2012advanced}.
%}

%\paragraph{Computation in 3D Cache}
%\textcolor{red}{
Exploitation of such CAM like structures for in-situ computing requires changes in the peripheral circuitry and user-space control. %over CAM structures.
Fortunately, in RRAM based memories, the control mechanism is relaxed in comparison to the DRAM counterparts, given that RRAM-based memories have no need to issue refresh (R), precharge (P), and activate (A) commands.
This makes it worth exploring the possibility of replacing these commands with the ones specifically used for software control over memory for in-situ computing.
In particular, we can design an API that communicates with the memory controller to perform application specific in-situ computation by repurposing the existing interface commands.
%The main advantage of following this approach is that RRAM based memories can be interfaced with DRAM compatible protocols, hence maintaining reusability of protocol.
%
%\textcolor{red}{
Notably, the use of CAM or RAM depends on the nature of user applications, hence a general solution with static allocation of RAM vs CAM is sub-optimal for the general case.
This paper explores reconfigurability of RRAM such that the same XAM array may be used in either CAM or RAM mode.
% depending on application requirements.%}

\paragraph{Design Challenges}
%\label{subsection:challenges}
%\textcolor{red}{
Last-level caches with 3D die-stacked memories incur significant write-bandwidth due to write backs from on-chip L3 and cache-line installation on cache-miss.
% from off-chip main memory. 
%For DRAM-based caches, this is not an issue. However, 
Non-volatile memories like RRAM usually have a limited write endurance, due to which some cells die after a specific number of writes to them (usually $10^8-10^{12}$~\cite{luo2019vlsit})
%lee2011fast, hsu2013self, chen2016doped}).
{This poses a significant challenge for employing RRAM-based caches, even more so because 3D stacked memories have a higher bandwidth than the conventional off-chip memory.}
This also poses security challenges where-in a malicious application can exploit frequent writes to the same block and render the memory unusable.
Monarch employs multiple techniques and solutions to keep lifetime at an acceptable level.
(1) Monarch introduces a new timing parameter, $t_{MWW}$, which imposes a timing constraint between successive writes to the same region of memory. This timing parameter cannot be enforced at fine granularity; therefore, Monarch introduces a set technique for viable implementation of the timing parameter.

(2) On top of the proposed timing constraint, Monarch employs a rotary technique at the same granularity to further enhance the lifetime, while maintaining a superior performance compared to an idealistic DRAM cache system.
(3) Monarch adopts optional write-through directly to the off-chip main memory, along with access no-allocate, thereby eliminating unnecessary writes to Monarch. % (Section \ref{section:lifetime}).
%This, coupled with $t_{MWW}$, helps limit the write bandwidth of Monarch to the same region and help keep lifetime of Monarch tolerable without significant performance loss.
%The aforementioned solutions help keep the lifetime of Monarch at acceptable levels with a tolerable decrease in performance.

\section{Overview}
\label{section:overview}
\begin{figure}[h!]
	\begin{center}
		\epsfig{file=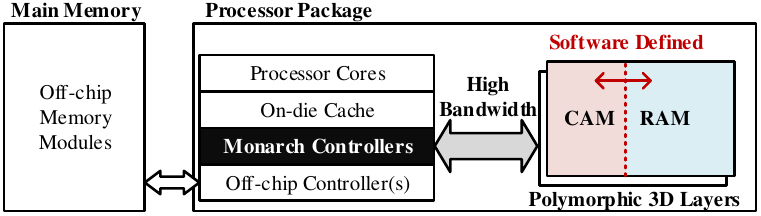, width = \columnwidth}
	\end{center}
	\caption{Overview of the proposed Monarch system.
		\label{figure:overview}}
\end{figure}

Figure~\ref{figure:overview} shows the overview of an example in-package Monarch. %in addition to the off-chip main memory within a multicore system.
Similar to the existing proposals for die-stacked DRAM~\cite{amdhbm}, a stack of 4-8 Monarch layers is connected to the processor die using a high bandwidth interface
% that includes through silicon vias (TSVs) and interposer-based routing techniques~
%\cite{tsv,interposer}.
The Monarch layers are divided into multiple independent \textit{vaults}, each of which has a separate controller. %set of dedicated fast links~\cite{tsv} for command, address, and data buses.
%A separate Monarch controller governs each vault to orchestrate data movement between the processor die and the 3D Monarch layers.
%All the controllers implement a set of commands and transactions defined by the JEDEC WideIO standard~
%\cite{jedechbm}.
% for high bandwidth memory systems.
%As it is explained later in Section~\ref{section:xwideio}, t
%The semantics of interfacing commands and their timing requirements are reevaluated based on the needs for Monarch operations.
%Within each Monarch vault, multiple banks share the vault interface to amortize the high cost of TSVs.
%Typically, 16 or more banks are employed to reach a high data bus utilization at the vault interface.
%Unlike the conventional 3D die-stacked memories, 
Monarch is capable of switching the functionality of each bank between random accessing (RAM) and content addressing (CAM) modes.
This novel capability enables the high-level software to choose between a hardware-managed cache policy and a scratchpad memory management for each Monarch vault.
Monarch follows a %boot-time 
similar approach to the one in Intel's Knights Landing~\cite{sodani2015knights} for setting the operational mode and {size of the RAM and CAM parts within each vault.} %\footnote{More details on the memory management and software control are provided in Section~\ref{subsection:os}.}

When a vault is configured for hardware-managed caching, the vault controller becomes responsible for automatic cache operations such as tag check, block installation, eviction, and replacement.
Therefore, a portion of the vault is turned into CAM for hardware-managed tag storage; while, the rest is used for storing cache data in the RAM mode.
The addressing space and operations of each hardware-managed vault remain invisible to the high-level software applications.
In contrast, a software-managed vault is visible to the user application through a specific address space for scratchpad operations.
Monarch extends this flexibility by allowing the scratchpad memory to be used in the RAM, CAM, or a mix of both modes.
%This high plasticity of Monarch is mainly due to employing the XAM array that efficiently performs both RAM and CAM operations with a minimal hardware overhead.

\section{XAM: Reconfigurable RAM/CAM}
\label{section:blocks}
\begin{figure}[h!]
	\begin{center}
		\epsfig{file=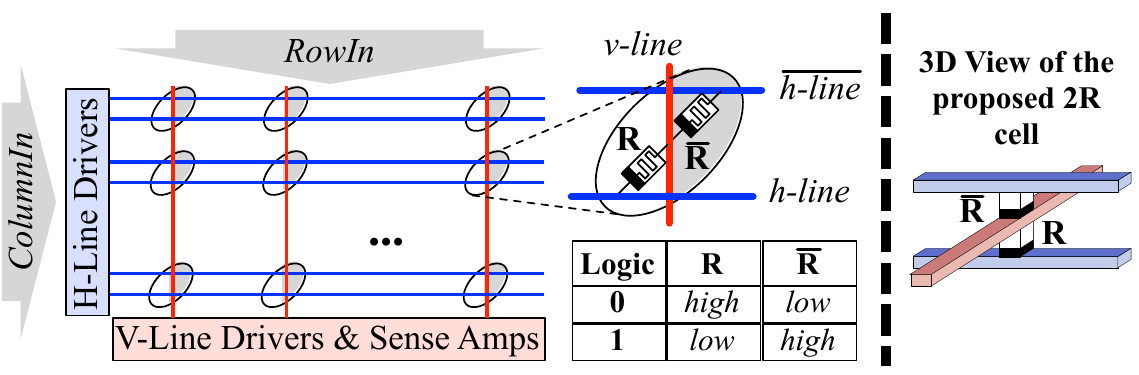, width = \columnwidth}
	\end{center}
	\caption{XAM array using differential cells.
		\label{figure:array}}
\end{figure}

The design of XAM (Figure~\ref{figure:array}) is based on the 2R cell structure proposed by the prior work, which enhances the read margin and alleviates the sensing error due to sneak currents and leakage in resistive crossbars.
%~\cite{chiu2014differential}
%We also adopt the sense-before-write scheme as proposed by the same work to eliminate the over-SET problem, while also improving lifetime.
% The 0,1 logic representation through this differential 2R is shown in the table in figure}~\ref{figure:array}
%
%  
%The key building block of Monarch is XAM that comprises a double-memristor crosspoint array for storing data bits in a differential format ($R$ and $\overline{R}$).
%\hl{
%As shown in Figure~\ref{figure:array}, XAM provides additional ColumnIn (H-Line) drivers per array to support both column writes and row writes.
%
% We also show parallel in-situ XNOR operation across cells in the same bitline, leading to a match vector being sensed at the V-line sense amplifiers, as explained in subsection}~\ref{subsection:search}. \hl{TODO: Talk to mahdi about parameters defined in the 2R paper.}
%
%
%Figure~\ref{figure:array} shows a XAM array built with the differential 2R cell. 
%Each 2R cell shares 2 horizontal lines ($h\_line$ and $\overline{h\_line}$) between other 2R cells in the same row, and shares a vertical line with 2R cells in the same column. 
Additionally, XAM supports both column and row writes.
The differential bit representation makes it capable of performing bit-wise \texttt{XNOR}, thereby parallel searching within the array.
Here, we explain the proposed operations for XAM.
%Therefore, in addition to the conventional read and write operations, in-situ bit comparison becomes a \textit{de facto} operation of each XAM array.
%Moreover, the crosspoint nature of the array together with a novel line activation mechanism enable arbitrary column and row writes to the array.
%

%with one $Output$ port and two possible input ports (i.e., $Row In$ and $Column In$).
%Each XAM cell comprises two memristive elements that are placed among three metal lines: two horizontal ($h\_line$ and $\overline{h\_line}$) and one vertical ($v\_line$).
%As shown in the 3D view, $\overline{h\_line}$ is laid above the plane of $v\_line$ in a perpendicular fashion, while memristors form the interconnects at every location a $h\_line$ passes above a $v\_line$. Similarly, $h\_lines$ is laid below $v\_lines$ in a perpendicular fashion. %The memristors between $v\_lines$ and $\overline{h\_lines}$ are laid in a similar fashion as the ones between $h\_lines$ and  $v\_lines$. 
%As the orientation of all memristors are vertically the same (figure~\ref{figure:array} 3D view), we expect a similar cell manufacturing as the one used for the Crossbar ReRAM Technology~\cite{liu2013130, Crossbar}.
%The horizontal lines are activated by a set of line drivers during a write or read operation.
%The vertical lines are connected to line drivers and sense amplifier to complete a read or write operation.

%
\begin{figure*}[h!]
	%\vspace{-2ex}
	\begin{center}
		\epsfig{file=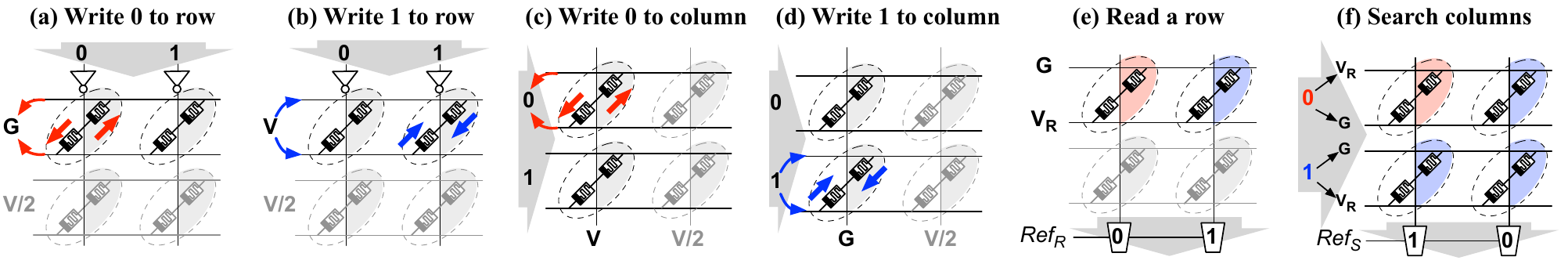, width = 2\columnwidth}
	\end{center}
	\caption{XAM operations: writing a row (a and b), writing a column (c and d), reading a row (e), and searching (f).
		\label{figure:operations}}
\end{figure*}
%

%\subsection{Differential Storage}
%
%
%Every XAM cell stores the bit value and its complement as $R$ and $\overline{R}$.
%Notice that $R$ is connected to $h\_line$ and $\overline{R}$ to $\overline{h\_line}$.
%The second terminal of both memristors are connected to $v\_line$.
%The low resistance state of the memristor represents \textbf{0}, while its high resistance state encodes \textbf{1}.

\subsection{XAM Write}
As shown in Figures~\ref{figure:operations}, writing data to XAM relies on the basics of programming bipolar memristive cells~\cite{pan2014recent}.
%A sufficiently high voltage (V) needs to be applied between the device terminals.
%The resistive state of the cell is determined by the direction of write current flowing through the device.
To make the figures easy to follow, every memristive element is shown with a black strip on one terminal.
Entering a write current from the black terminal sets the resistive state to low; otherwise, the cell is programmed to a high resistive state.
Similar to the prior work~\cite{xu2015overcoming}, 
XAM follows a two-step write process for writing \textbf{0}s and \textbf{1}s.
However, special considerations are necessary to enable row- and column-wise wirtes.
%However, through the introduction of reconfigurability at boot-time, the XAM arrays operates either in the row mode or column mode at a given time.
%Notice that a switching between the write dimensions leads into flipping the direction of write current through the memristive elements; therefore, the data bits stored through column and row become inconsistent.
%XAM addresses this issue by inverting the data bits before a row write, while the bits are stored without inversion through a column write.

\subsubsection{Writing a XAM Row}
For writing a row, a multi-bit input is inverted and fed in through the column drivers at the top edge (Figures~\ref{figure:operations} (a) and (b)).
All the vertical line corresponding to input \textbf{0}s are connected to $V$, while the remaining vertical lines (representing \textbf{1}s) are set to ground ($G$).
%During a row write, o
Only one row is active and all the other rows are kept inactive by connecting them ro $V/2$.
%(This is similar to the unselected rows in the conventional crosspoints~\cite{xu2015overcoming}.)
To write a \textbf{0}, the horizontal lines of the active row are first connected to ground ($G$), which creates a potential drop and a current flow as shown in Figure~\ref{figure:operations}(a).
This current flow corresponds to writing a \textbf{0} into the selected XAM cells whose vertical line is connected to $V$.
All the other cells of the array remain unchanged.
The second step begins by switching the voltage level of active horizontal lines from $G$ to $V$, which results in stopping the flow of current for writing \textbf{0}s.
Instead, new currents for writing \textbf{1}s start to flow (Figure~\ref{figure:operations}(b)).
%In summary, only \textbf{1}s are written to the active row during the second step.

\subsubsection{Writing a XAM Column}
%The two steps of writing a column to the XAM array are shown in Figures~\ref{figure:operations}(c) and (d).
The data to be written is fed in through the row driver onto the horizontal lines (Figures~\ref{figure:operations}(c) and (d)).
Similarly, \textbf{0}s are written first followed by \textbf{1}s. % the second step.
Initially, both horizontal lines of each row are connected to either $G$ or $V$ with respect to the bit value being stored in the row.
%(Unlike writing a row, now the horizontal lines are set to fixed voltages while the vertical lines switch.)
%During a column write, o
One column is activated while the rest are kept inactive by setting their vertical lines to $V/2$.
First, the active column is connected to $V$ for writing any input \textbf{0}s.
%As shown in Figure~\ref{figure:operations}(c), 
Notably, writing a \textbf{0} row-wise and column-wise produce the same cell state.
Second, a voltage switch from $V$ to $G$ is necessary to stop writing \textbf{0}s and to begin storing \textbf{1}s (Figure~\ref{figure:operations}(d)).

\subsection{XAM Read and Search}
\label{subsection:search}
%Every XAM array can be configured for RAM or CAM operations through external commands generated by the Monarch controller. %\footnote{Details of the Monarch interface are provided in Section~\ref{section:xwideio}.}
Configuring a XAM array for reading or searching requires (1) a XAM datapath that supports communicating address/data bits and (2) a proper reference for sensing bit values.

\subsubsection{Reading a XAM Row}
Figure~\ref{figure:operations}(e) shows the process of reading a XAM row.
First, the row is activated for reading by connecting its $h\_line$ and $\overline{h\_line}$ to a read voltage ($V_R$) and ground, respectively.
%This particular activation of the row ensures connecting the low resistive element of each stored \textbf{0} to $G$ and that of each \textbf{1} to $V_R$.
As a result, each XAM cell becomes a voltage divider between its horizontal lines to develop a potential on its vertical line.
The amount of this voltage is determined by $\frac{\overline{R}}{R+\overline{R}}\times{V_R}$. %, which is a function of the stored bit.
Intuitively, a \textbf{0} ($\overline{R}=low$) corresponds to a near $G$ potential and a \textbf{1} ($R=high$) develops a near $V_R$ voltage.
%At the bottom of each horizontal line, a
A sensing circuit is employed to compare this potential against a read reference ($Ref_R$) for producing a bit of the output.
%~\cite{geiger.book90} 
We set $Ref_R$ to $V_R/2$ for reads.
%
%\begin{figure}[h!]
%	%\vspace{-2ex}
%	\begin{center}
%		\epsfig{file=figures/read.pdf, width = \columnwidth}
%	\end{center}
%	\vspace{-2ex}
%	\caption{Illustration of a read (a) and a search(b).
%		\label{figure:read}}
%	%\vspace{-2ex}
%\end{figure}

\subsubsection{Searching among XAM Columns}
A XAM search can be viewed as an extended read operation, where the horizontal lines are activated based on the contents of an input key and the sensing reference is recomputed to differentiate between a column match and mismatch (Figure~\ref{figure:operations}(f)).
First, the input key is applied to the horizontal line drivers.
A \textbf{0} results in connecting $h\_line$ and $\overline{h\_line}$ respectively to $G$ and $V_R$; while, a \textbf{1} creates the opposite line voltages.
This complementary application of the key bits to the horizontal lines results in two possibilities for each cell:
(1) the low resistive element of the cell is connected to ground that corresponds to a bit \textbf{mismatch}; and (2) the low resistive element is connected to $V_R$ that reads to a bit \textbf{match}.
Then, the sensing circuits convert the line voltages to binary values.
If all the cells within a column are match, the vertical line voltage remains near $V_R$; therefore, the sensing circuit produces a \textbf{1} at the output, which indicates a column match.
Otherwise, each bit mismatch pulls down its line voltage such that even a single bit mismatch can drop the voltage below $Ref_S$.
Therefore, the column will be represented by a \textbf{0} at the sensor's output.
For a XAM array with $N$ rows, $Ref_S$ should be set to a voltage between all match ($\frac{H}{L+H}\times{V_R}$) and single mismatch ($\frac{L+(N-1)H}{L+H}\times{V_R}$); where, $H$ and $L$ correspond to the high and low resistances of the memristive elements.

\section{Choice of Technology for RAM/CAM}
\label{section:technology}
%Since we propose an in-package 3D-stacked reconfigurable memory, it is important to know why novel XAM arrays should be preferred over other technologies for the case of software reconfigurable memory. To this end, we do 
This section compares the latency, energy, and area overheads of a 32KB reconfigurable RAM/CAM block using the CMOS, DRAM, and RRAM technologies (Table~\ref{table:tech_comparison}).

\paragraph{Latency} The CMOS unit for SRAM+SCAM has the best read latency, followed by 1R and XAM. %. XAM has a slightly worse read latency than 1R.
SRAM has a $10\times$ and $100\times$ better write latency than DRAM and RRAM.
The search latency of SRAM, DRAM, and 1R are much higher than the others due to the fact that they support serial search rather than parallel. %there is no support for parCAM reads, and so to force CAM reads, the controller can only access one bit of information per CAM cycle. CMOS-CAM
Instead, SCAM provides the shortest lookup time, followed by XAM and 2T2R.

\paragraph{Energy} SRAM and SRAM+SCAM exhibit the best read energy, followed by 1R and XAM. The search energy of XAM and 2T2R are low due to their efficient in-situ XNOR operations.
SCAM requires a low write energy but a significant search energy due to its cell complexity.
An RRAM write require magnitudes higher energy than SRAM and DRAM.
%SCAM exhibits the best write energy.

\paragraph{Area} The 1R array consumes the least area, which is similar to that of XAM. The area of a XAM array is about $10\times$ smaller than the SRAM+SCAM. %, which is not reconfigurable but nevertheless includes both CAM/RAM operation.
\begin{table}[h!]
	\vspace{-3ex}
	\caption{A 32KB building block in various technologies~\cite{balasubramonian2017cacti,2012nvsim}.\label{table:tech_comparison}}
	\vspace{-2ex}
	\scalebox{0.75}{
		\begin{tabular}{l|ccc|ccc|c|}
			\cline{2-8}
			& \multicolumn{3}{c|}{\textbf{Latency (ns)}}       & \multicolumn{3}{c|}{\textbf{Energy (nJ)}}        & \multicolumn{1}{l|}{\textbf{}} \\ \cline{2-7}
			& \textbf{Read} & \textbf{Write} & \textbf{Search} & \textbf{Read} & \textbf{Write} & \textbf{Search} & \textbf{Area (mm2)}            \\ \hline
			\multicolumn{1}{|l|}{\textbf{SRAM}}      & 0.2334        & 0.1892         & 14.9395         & 0.015         & 0.0196         & 0.9627          & 0.0331                         \\
			\multicolumn{1}{|l|}{\textbf{SCAM}}      & 32.2385       & 0.2167         & 0.5037          & 0.2329        & 0.0139         & 0.1273          & 0.111                          \\
			\multicolumn{1}{|l|}{\textbf{SRAM+SCAM}} & 0.2334        & 0.2167         & 0.5037          & 0.015         & 0.0335         & 0.1273          & 0.144                          \\
			\multicolumn{1}{|l|}{\textbf{DRAM}}      & 2.5945        & 2.1874         & 166.0499        & 0.0657        & 0.058          & 4.4544          & 0.0169                         \\
			\multicolumn{1}{|l|}{\textbf{1R RAM}}    & 1.654         & 20.258         & 105.856         & 0.0214        & 0.325          & 1.623           & 0.0104                         \\
			\multicolumn{1}{|l|}{\textbf{2T2R CAM}}  & 122.048       & 20.825         & 3.36            & 2.7156        & 1.29           & 0.0472          & 0.0153                         \\
			\multicolumn{1}{|l|}{\textbf{1R+2T2R}}   & 1.654         & 20.825         & 3.36            & 0.0214        & 1.61           & 0.0472          & 0.0258                         \\
			\multicolumn{1}{|l|}{\textbf{2R XAM}}    & 1.7734        & 20.323         & 3.2264          & 0.0215        & 0.652          & 0.0263          & 0.0124                         \\ \hline
		\end{tabular}
	}
	\vspace{-2ex}
\end{table}

%,chiu2014differential,5746281,

%\hl{TODO: References for each of the technologies}

%\hl{TODO: FGDRAM story, and explain building 3D stacked memory}

Overall, 1R, SRAM+SCAM, and XAM could give configurations optimized for the specific metrics mentioned above.
DRAM is a default baseline considering its widespread use in the existing off-chip and on-chip memories.
Therefore, we select three technologies to build the 3D baseline system: 1R RRAM, CMOS SRAM+SCAM, and DRAM. 
Note that XAM process is CMOS compatible, thereby manufacturable on the processor die.
But its capacity would be very less to capture the complete behavior of big data applications.
On the other hand, off-chip integration of Monarch would encounter a limited bandwidth making it a poor choice for caching.
%, which is an important requirement for web servers etc (TODO: References). Keeping this in mind, Monarch would be more useful used 
Therefore, we consider Monarch as an in-package high-bandwidth memory flexibly used for cache, associate search, and scratchpad memory.
%reconfigurable , because that guarantees gigascale-storage while still ensuring high-bandwidth for caching performance.

\paragraph{Stack Model}
Considering a 4GB DRAM stack with 8 layers divided into 8 vaults, a total of 64 segments each sized 64MB are necessary.
We consider the same configuration for all the baselines using segments that consume the same area across all the technologies.
This provides an iso-area comparison among the systems, with memory capacities of 8GB (1R RRAM), 8GB (Monarch), and 73MB (CMOS).
%The stack memory capacities 
% using the same to build iso-area each) baseline, the RRAM baseline for the same area and number of layers has a capacity of around 8GB (64 segments of size 128MB each). SRAM for the same area comes to a size of 320MB (64 segments of size 5MB each). A CMOS-CAM/RAM approach, which ensures flexible CAM or RAM performance of separate building blocks depending on the needs, has a capacity of 73.28MB for the same area and number of layers. Moreover, we use low standby-power versions of the transistors for the CMOS-CAM/RAM baseline, giving us a good approximation of new 3D-stacked SRAM products such as the AMD Ryzen~\cite{ryzen}. 

\section{Monarch Organization}
\label{section:xcache}
Monarch follows the HBM2 3D organization proposed by the prior work on Fine-Grained DRAM~\cite{o2017fine}.
%\hl{This organization helps reduce data movement energy by partitioning the XAM arrays into independent grains ( each with adjacent local I/O), and also helps maximizing internal bandwidth by exploiting local I/O (as explained in }\textit{Diagonal Set Arrangement }\hl{below). We also perform RRAM 3D-stacking similiar to the HBM2 configuration in Fine-Graned DRAM.} 
However, a novel datapath is used to support RAM and CAM arbitration.
%operations are necessary.
%
%\subsection{Physical Layout}
%\label{section:layout}
%Figure~\ref{figure:layout}(a) illustrates the physical layout of a 4-layer Monarch stack.
%All the layers as a whole are divided into multiple vaults.
%A TSV stripe crossing the layers is employed to connect the Monarch vaults and the processor die.
%Based on an Monarch interfacing protocol, a portion of the TSV stripe is dedicated to each vault for communicating the address, command, and data bits.
%%
%Each vault includes four segments that are placed in different layers.
%Every segment is further divided into 8 banks that share the same TSVs with all the other banks in the vault.
%Each Monarch bank is designed to operate in the CAM or RAM mode independently.
XAM enables such flexibility at low cost by leveraging a toggle-based mechanism to set the operational mode of each bank. %\footnote{Section~\ref{section:control} provides more details on the Monarch control.}

\subsection{Monarch Physical Layout}
\label{section:layout}
\begin{figure}[h!]
	\begin{center}
		\epsfig{file=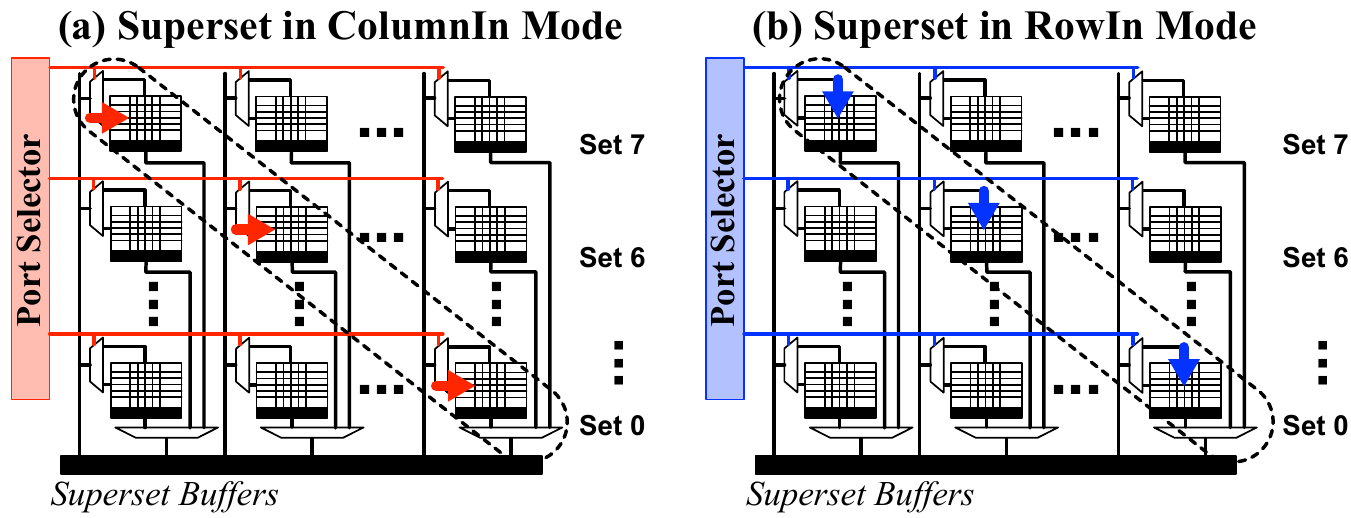, width = \columnwidth}
	\end{center}
	\caption{Diagonal selection for ColumnIn (a) RowIn (b) modes.
		\label{figure:layout}}
\end{figure}

Within each Monarch bank, XAM arrays are physically grouped into supersets that share the same H-trees for data and address.
Each superset consists of data/mask/key buffers, a port selector, and multiple XAM arrays (Figure~\ref{figure:layout}).

\paragraph{\textbf{Diagonal Set Arrangement}}
A key challenge for designing supersets is the increasing complexity of interconnects for supporting CAM and RAM operations.
% in both RowIn and ColumnIn modes.
We alleviate this complexity through (1) sharing the address and data interconnects at the bank and vault levels and (2) accessing the superset arrays diagonally.
As shown in Figure~\ref{figure:layout}, sets are organized diagonally in an $8\times8$ superset: each subarray at row $i$ and column $j$ belongs to set {$k$ ($=(j-i)\%{8}$).}
For every access, only 8 subarrays of a set are selected.
The port selector's job is to enable the path for sending inputs to the right port of the subarrays.
Figure~\ref{figure:layout}(a) shows how the column ports for set 0 are selected; whereas, row ports are enabled in  Figure~\ref{figure:layout}(b).
As the diagonal layout of the sets is fixed, the port selector only requires a latch for holding a mode flag and a 3-to-8 decoder for selecting the corresponding XAM arrays.
%Similar to activating the CAM/RAM modes in banks, we employ a toggle-base mechanism for switching between the two input ports.

\subsection{Monarch Interface}
\label{section:xwideio}
\begin{figure}[h!]
	\begin{center}
		\epsfig{file=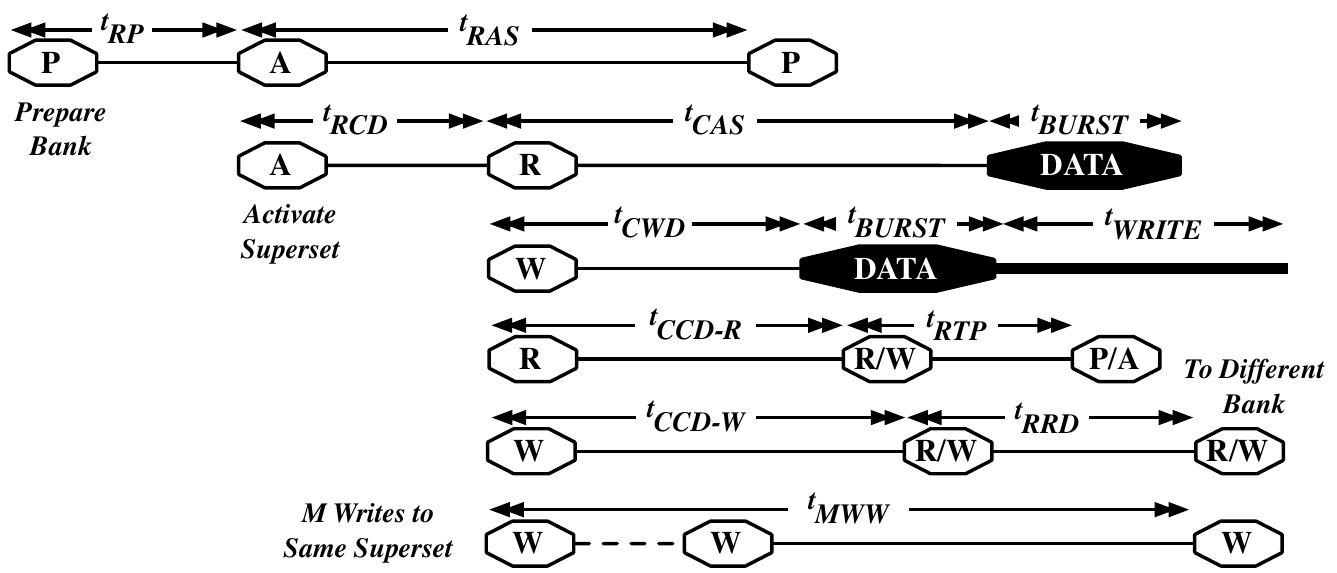, width = 1\columnwidth}
	\end{center}
	\caption{Timing details of the Monarch interface.
		\label{figure:protocol}}
\end{figure}

%
%\hl{Major TODO}
Monarch is based on the proposed resistive XAM technology that provides a slower write speed and a comparable read speed to DRAM.
One effective technique to reduce the write latency in RRAM is multi-banking that has been frequently suggested by prior work.%~\cite{mittal2018survey}.
Following the same logic, Monarch employs a high bandwidth interface that allows the processor to access a large number of parallel banks across multiple vaults.
The Monarch interface is designed after the JEDEC high-bandwidth DRAM~\cite{jedechbm}.
DRAM necessitates a set of parameters and commands---i.e., refresh, precharge, activate, read, and write---to be included in the standard.
As the DRAM command are not sufficient for XAM operations, we redefine the semantics of parameters as listed in Table~\ref{table:timing}.
Therefore, we map the existing commands to XAM-semantics and derive XAM-specific values for the timing parameters.
Figure~\ref{figure:protocol} illustrates a summary of key timing parameters proposed for the Monarch interfacing protocol.

\begin{table}[h!]
	\caption{Key timing parameters used in Monarch.\label{table:timing}}
	\centerline{
		{\scriptsize\setlength\tabcolsep{3pt}
			\begin{tabular}{|r|l|}
				\hline
				%\textbf{Parameter} 
				& \textbf{Description}\\ % & \textbf{Time} \\
				\hline
				\hline
				$t_{RP}$ & bank preparation time including setup delay for $Ref$\\%~\cite{ker2006design}\\
				$t_{RAS}$ & superset activation time (mainly datapath activation delay)\\
				$t_{RCD}$ & same as $t_{RAS}$\\
				$t_{CAS}$ & the time to complete a read or search (depending on the\\& bank mode) and transferring sensing results to the vault interface\\
				$t_{CWD}$ & the time for sending address and command bit to the TSV stripe\\
				$t_{CCD-R}$ & read cycle time (max of the interconnects  delay and sensing)\\
				$t_{WRITE}$ & the time to complete the proposed 2-step data write\\
				$t_{CCD-W}$ & write cycle time (max of the interconnect delay and $t_{WRITE}$)\\
				$t_{RTP}$ & the time to transfer data from the TSV stripe to a set\\
				$t_{RRD}$ & same as $t_{RTP}$\\
				$t_{BURST}$ & the time to burst data in the TSVs and silicon interposer\\
				$t_{MWW}$ & the time window for \textit{M} writes\\
				\hline
		\end{tabular}}
	}
\end{table}

\paragraph{\textbf{Preparing a Bank}}
%As shown in the figure, 
Monarch introduces a \textit{prepare} command (\textbf{P}) as a replacement for the DRAM precharge.
During a bank \textit{prepare}, the Monarch controller toggles the mode of a bank between CAM and RAM.
The initial mode of all Monarch banks is set to RAM.
In DRAM, a precharge is necessary in the case of a row conflict---i.e., the row buffer contains a different row than the one is to be accessed.
DRAM row conflicts may happen frequently during the execution of a program.
Whereas, a bank \textit{prepare} is only necessary if the controller decides to change the read mode of all the XAM arrays inside the bank.
Upon receiving a \textit{prepare} command, the bank controller makes an internal state transition by switching the sensing reference between $Ref_R$ and $Ref_S$ using bank-level voltage level converters~\cite{geiger1990vlsi}.
Moreover, the supersets inside the bank will be configured for upcoming search ($Ref_S$) or read ($Ref_R$) operations.
The default read mode of each bank is read, which makes it possible for the Monarch controller to track the mode of all banks.

\paragraph{\textbf{Activating a Superset}}
Instead of a DRAM row activation, Monarch may issue an \textit{activate} command (\textbf{A}) to a superset before bursting data on the interface.
The main purpose of the new \textit{activate} command is to change the internal datapath of a superset prior to writing data in the \textit{RowIn} and \textit{ColumnIn} modes.
The DRAM activates, however, are necessary for accessing the missing rows in the row buffer.
To keep the controller simple, issuing an activate command toggles the operational mode of the port selector between the column and row access.
A bit flag for each superset is adequate to track the state of all supersets at the Monarch controller.

%\paragraph{Bursting Data}
%Similar to the DRAM interface, the \textit{read} (\textbf{R}) and \textit{write} (\textbf{W}) commands are used to initiate a data burst between the Monarch controller and the data buffer in a superset.
%Regardless of the mode, a \textit{read} command activates all the XAM arrays within a selected set to produce data through local sensing.
%Next, the local data is collected in the superset buffer and sent to the Monarch controller via the on-chip interconnects and the TSVs.
%For writing data to a superset, the data is first transferred to the data buffer at the superset.
%Then, the two-step process for writing resistive cells is applied to write the data row- or column-wise.
%%To alleviate the high cost of resistive writes, we leverage the differential writes~\cite{zhou2009durable} at the XAM arrays.
%
%%Notice that Monarch does not make any use of the refresh command.
%%With these modifications, Monarch keeps the physical structure of the interface identical to that of high bandwidth DRAM.
%%However, the Monarch vault controller will become significantly different from a DRAM controller.

\paragraph{\textbf{Fine-grained XAM Access}}
In addition to data buffer, each superset employs a mask and a key buffer for accessing the selected set in the CAM mode.
Prior to a search, the keys and bit-masks are sent from the Monarch controller to the superset.
We use write command for key/mask transfer.
To distinguish between a CAM write and a key/mask write, the superset does not allow writing an input block to the XAM arrays if the mode is set to RowIn for CAM.
%(The contents of a CAM configured array may only be updated in the ColumnIn mode.)
Instead, a block in the RowIn mode is only written to the mask or key registers if the row address is odd or even, respectively.
This enables searching keys with different lengths and arbitrary alignments within arrays.
The mask register is also used for partial updates to the data in the ColumnIn mode.
This capability is useful for updating the dirty bits of cache tags in one access (Section~\ref{section:control}).

\paragraph{\textbf{Constraining Block Writes}}
To ensure a lower bound lifetime for Monarch, we introduce a new timing parameter ($t_{MWW}$) that limits the number of writes per block within a time window.
%constraints multiple window writes per superset.
%
%
%$t_{MWW}$ timing parameter acts as a user knob for controlling the worst case lifetime we ensure for Monarch. 
%Assume $n_{w}$ corresponds to the 
Given a write endurance ($n_{W}$) for the XAM cells, the minimum lifetime ($T_{Life}$) is computed by $T_{Life}=n_W{\times}t_W$, where $t_W$ is the time window during which a single write per block is allowed.
To allow $M$ writes per window while guaranteeing the same lifetime, a wider time window is necessary, which is computed by
$t_{MWW}=\frac{M{\times}T_{Life}}{n_W}$.
%The window 
%, while a , the guarantee a user needs, and $w_{WIN}$ corresponds to the number of writes allowed in a window,  then $LT = n_{w} \times \frac{t_{MWW}}{w_{WIN}}$. Given this formula, user can calculate the timing parameter required for a specific write endurance, writes per window and lifetime.
%
For example, to ensure a 3-year lifetime ($94.6\times10^6$ seconds) with a cell endurance of $10^{8}$ writes,  $t_{MWW}$ has to be $0.94M$ seconds.
%can be $\frac{1{\times}3{\times}365{\times}24{\times}3600}{10^8}$ for $M=1$ before being worn out and the user expected lifetime is 3 years, for $t_{4WW}$ (A maximum of 4 writes to a single block within a window), the value would be $\frac{3}{10^{8}} \times 356 \times 24 \times 60 \times 60 \times 4$, resulting in $t_{4WW} = 3.691$ seconds. For a single write window, that value would be 0.922 seconds
Indeed, naively enforcing such timing constraint is impractical due to high complexity and performance overheads.
Section~\ref{section:lifetime} explains how Monarch implements $t_{MWW}$.

\section{Monarch Control}
\label{section:control}
Monarch %is designed to 
supports
% runtime reconfiguration of an in-package memory for 
three modes: flat-RAM, flat-CAM, and cache.
The flat modes are managed by software through their visible address spaces at the application level.
However, the cache mode is completely hardware-managed at the Monarch controller.
%Similar to the Intel's KNL~\cite{sodani2015knights}, the proposed memory configuration is only allowed for vaults at boot time.
For example, an eight-vault Monarch may be configured to 2 flat-CAM, 2 flat-RAM, and 4 cache vaults.
%Each Monarch vault controller is responsible for translating a stream of memory requests into Monarch commands while enforcing the timing parameters explained in the previous section.
%Moreover, the requests may be serviced differently according to the vault mode.

\paragraph{\textbf{Flat-RAM Control}}
A flat-RAM vault is supposed to service only two types of requests (i.e., read and write) generated by software.
%As a result, the simplest mode of Monarch controller is for orchestrating data within a flat-RAM vault; where, a set of 
Multiple counters are used to track the required delays between pairs of commands before sending each command to the Monarch layers.
Moreover, the controller needs to ensure that the target bank/superset is set to RowIn RAM. %prior to every read/write access; otherwise, a prepare and an activate may be required.

\begin{figure}[h!]
	\begin{center}
		\epsfig{file=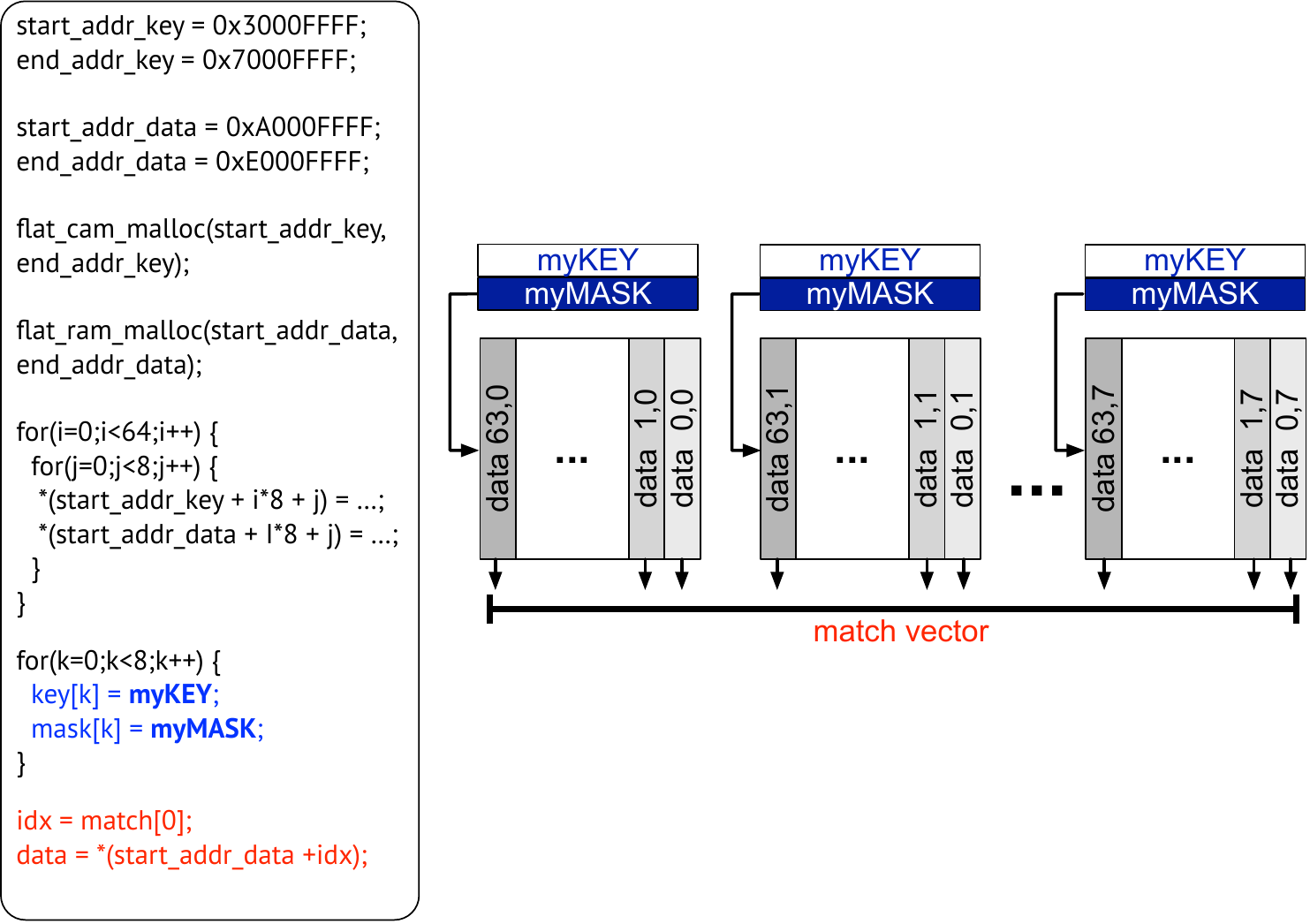, width = \columnwidth}
	\end{center}
	\caption{Searching a set using key and mask buffers.
		\label{figure:search}}
\end{figure}

\paragraph{\textbf{Flat-CAM Control}}
%As compared with the flat-RAM control, additional 
Extra components are necessary to enable %controlling a vault in 
the flat-CAM mode.
First, the controller needs to recognize four different types of software generated requests: data write, search, key/mask write, and data read.
Second, the controller may require transitioning the target Monarch bank/superset to different modes prior to serving a request.
A data write may be carried out only if the target bank/superset is in the ColumnIn CAM mode.
The software is expected to populate the XAM arrays using data writes prior to a search.
Figure~\ref{figure:search} shows an illustrative example code for operation of CAM through a simple key-value store. The key and data values are initially written into the CAM and RAM scratchpads (a single set each in this example), which is followed by updating the key/mask registers, searching the set, and accessing data based on the value returned by the match pointer.

Four pointers are employed for accessing the elements of data, key, mask, and match.
%(As will be explained later in this section, a software library supporting Monarch provides the pointers.)
For simplicity, assume that the data is a 2D array of a 64-bit type. %; therefore, each data element is stored in a subarray column.
Each block of 8 data elements is stored across 8 subarrays.
A total of 64 blocks are written in a single set. \textit{flat\_RAM\_malloc} and \textit{flat\_CAM\_malloc} are the proposed function calls for allocating memory within the RAM and CAM scratchpad regions respectively, explained in detail later in this section.
Next, the key/mask values are updated through 8 normal writes in the code.
However, any accesses to the key/mask pointers are recognized by the flat-CAM controller for special service.
First, the key/mask pointers are mapped onto two global registers in the vault controller, which are used for any upcoming search operations.
Second, the contents of these registers are sent to the target supersets prior to every search.
To eliminate any unnecessary key/mask updates, the controller tracks the recently updated supersets.
Each key/mask write may only be carried out in the RowIn CAM mode; therefore, parepare and activate commands may be necessary to toggle the bank/superset mode.
%As shown in the figure, a 
A search may be initiated by the software through a normal read to the match pointer. 
Similarly, the match pointer is mapped to a special register at the Monarch controller. 
On every read to this register, the memory address is used to determine the target Monarch superset. 
On a search, the match register is updated with the matching index of the search, and it gets reset to NULL if there is no match in the specific superset. The value stored in the match register can be accessed by the program (Figure~\ref{figure:search}).
The controller will issue a search to the target superset if the results of previous search is not present in the match register and the key/mask registers have not yet been updated.
When the previous search result is not present in the match register, a key/mask update is necessary if the superset does not have the latest key/mask values.
Prepare and activate commands may be necessary before a read in the ColumnIn CAM mode to perform a search at the target Monarch set.

As the key/mask registers are shared by multiple sets within each superset, searching consecutive sets per superset do not incur unnecessary key/mask updates.
Moreover, partial search and variable key length is supported through masking.
For example, setting myMASK to \textit{0x0FF00} results in searching among the second byte of data elements.
\footnote{When we have to read the actual keys stored inside the CAM arrays, the controller issues read commands in row mode.}

\begin{figure}[h!]
	\begin{center}
		\epsfig{file=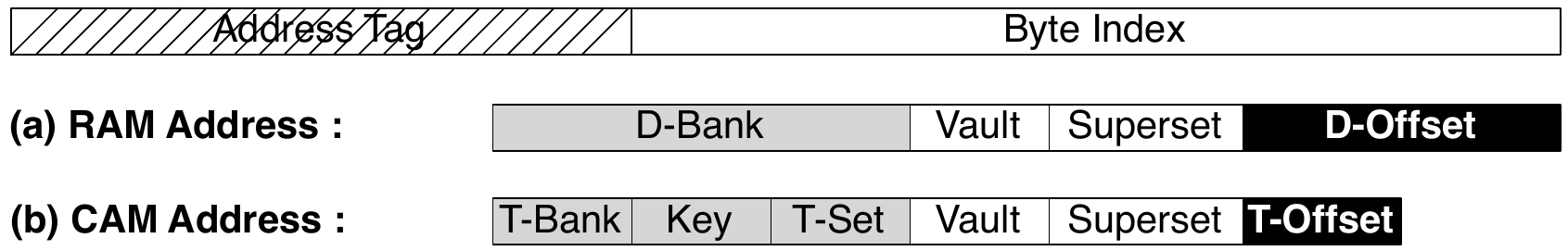, width = \columnwidth}
	\end{center}
	\caption{Coordinated mapping of the RAM and CAM addresses for requests.
		\label{figure:address}}
\end{figure}

\paragraph{\textbf{Cache Control}}
%The most sophisticated control policy is used for hardware managed caching, where 
In the cache mode, the vault requires both CAM and RAM banks simultaneously.
For example, a vault with 64 banks may be partitioned into 30 RAM and 2 CAM banks to support 64B cache blocks tagged with 30 address bits, 1 valid bit, and a dirty flag.
%As soon as a vault is configured for cache control, the 
The RAM and CAM banks are assigned for data and tag storage respectively.
Upon receiving a request, the vault controller follows certain steps to access %obtain the data from the 
Monarch or forward the request to memory.
As the data and tags are physically decoupled across banks, we need a coordinated address mapping that corresponds the tag and data locations within each vault.
First, we consider a 512-way set associative cache due to the available search bandwidth across all the set columns for one access.
For every 512 tags maintained in a set, 512 data blocks are stored in a superset.
Therefore, every CAM set corresponds to a RAM superset.
Second, every data block requires a tag. Hence, the number of tag locations must be greater than or equal to the number of stored data blocks.
For example, $\frac{1}{16}$ of the CAM partition in a cache vault with 32b tags and 32 banks is left unused. %\footnote{Which is used for wear-leveling purposes (Section~\ref{subsection:lifetime}).}.
As shown in Figure~\ref{figure:address}, every memory address is translated to RAM and CAM addresses that have the same vault and superset IDs.
The bank ID of the RAM address is partially used for computing the set, key and bank IDs of the CAM address.
Notably, each array column stores two 32-bit tags.
The key ID is used to determine which tag of a row to access during a search.
Every tag lookup requires searching for a key/mask combination.
We choose more significant bits of the CAM address for the key ID to reduce the frequency of mask updates.

For every cache lookup %involves search in the CAM banks.
%First 
(1) the key (and if necessary the mask) values are updated
and (2)
%Then, 
a search will be issued by the controller.
Based on the search outcome, a hit or miss will be detected at the controller.
On a hit, the corresponding data block will be accessed in the RAM part; otherwise, the request will be forwarded to the main memory.
To install a cache block in Monarch, a data read in RAM mode to the CAM part  is necessary to check the valid bits of all the 512 tags in a set.
Upon finding an invalid tag, the location will be selected for installing the new data and tag.
Otherwise, another row read to the dirty bit position of the tags, determines if a clean block is available for eviction.

\paragraph{\textbf{OS Support}}
\label{subsection:os}
%Memory channel/vault configuration has been considered for improving the efficiency of resource utilization.
%For example,  
Intel's KNL employs the memkind library extension~\cite{memkind} that, which allows the applications to call custom memory allocation (\textit{malloc}) functions.
\textit{HBM\_malloc} is such an example where the memory is allocated in the in-package HBM rather than main memory.
Monarch extends the principle of custom memory allocation from the memkind library to add two new function calls (\textit{flat\_RAM\_malloc} and \textit{flat\_CAM\_malloc}) forthat allow us to independently allocatinge memory for requests in the RAM and CAM vaults respectively.
%In addition to allocating memory in different Monarch vaults, the 
Also, the extended library returns pointers to the match and key/mask registers defined inside the vault controllers.
%As a result, the proposed Monarch hardware may be employed by user applications.
%
%One key limitation of the conventional software hardware abstraction is the plain virtual memory that treats all data types similarly during page allocation and address translation.To overcome this limitation, it is imperative to have methods that makes virtual pages visible to the memory hierarchy. This will allow for user-defined memory mapping, where user can control where in memory specific data is placed.
%
%Such a method should support appending information to each memory allocation request, which will enable to application to establish some instruction to the OS to allocate memory for that request in a specific way.
%
%\paragraph{Memkind} 
%
%From the Hardware perspective, Monarch configuration happens at a channel level granularity. Independent Channels can be configured in either CAM mode/RAM mode or Cache mode. Allocation into RAM/CAM mode happens through the aforementioned extension, while Cache mode is purely hardware defined.

\section{Lifetime of Monarch}
\label{section:lifetime}
The lifetime of memristive cells are mainly impacted by write operations. 
The cell write endurance is a property of the device material, which is influenced by the write current and time.
Recent device experiments on RRAM have shown a write endurance of over $10^8$ up to $10^{12}$~\cite{luo2019vlsit}.
%~\cite{kim2011bi, lee2010evidence}
We consider a cell endurance of $10^8$ for the evaluations.
Other parameters impacting lifetime are the number of cells (memory size) and write traffic.
Monarch employs three techniques to lower write traffic at low cost.

\paragraph{Tracking Writes}
Tracking block writes with $t_{MWW}$ may be costly and impractical.
Consider 4-byte storage per block to track the last writes, a total of 256MB memory would be necessary to track the blocks.
Instead, we exploit the specific properties of Monarch to reduce this cost significantly.
First, the counters are maintained at the superset granularity, while the writes within each superset are evenly distributed across blocks (will be explained later).
Second, the counters are stored in the main memory while a TLB-like approach is employed for on-chip buffering.
Due to the high locality of memory accesses, we observe a negligible performance impact by this mechanism.
% the counters on-chip
%roperties of in-package memory and share the counters for a of storage per superset, resulting in total storage overhead of 128KB for a 4GB system (refer to Table ~\ref{table:system} for Monarch configuration). Having it per block for the same configuration would result in storage overhead of 64MB, which is infeasible.} 

%\textcolor{red}{$t_{MWW}$ timing parameter acts as a user knob for controlling the worst case lifetime we ensure for Monarch. Say $n_{w}$ corresponds to the write endurance of each cell in Monarch, $LT$ is the minimum lifetime guarantee a user needs, and $w_{WIN}$ corresponds to the number of writes allowed in a window,  then $LT = n_{w} \times \frac{t_{MWW}}{w_{WIN}}$. Given this formula, user can calculate the timing parameter required for a specific write endurance, writes per window and lifetime.}

%\textcolor{red}{Assuming each memristive cell in Monarch can withstand $10^{8}$ before being worn out and the user expected lifetime is 3 years, for $t_{4WW}$ (A maximum of 4 writes to a single block within a window), the value would be $\frac{3}{10^{8}} \times 356 \times 24 \times 60 \times 60 \times 4$, resulting in $t_{4WW} = 3.691$ seconds. For a single write window, that value would be 0.922 seconds}

By enforcing $t_{MWW}$ at the 512-block supersets, a new write will not be allowed for $t_{MWW}$ if the superset writes exceed $512M$.
Monarch follows a strict blocking policy for such write to ensure the target lifetime.
Hence, the programmer's responsibility is to map read-mostly data to Monarch in the flat mode.
None of our evaluated applications in the flat mode exceeds this timing requirement.
In the cache mode, however, the superset will be locked for $t_{MWW}$ and all its accesses are forwarded to the main memory.
% will be used alternatively.
% exceeding 
% \textcolor{red}{In case we have a misprediction, in the case of a single write window, the superset can't be written into for the next 0.922 seconds, leading to a major performance impact. Having multiple writes within a window helps mitigate the performance effect of a misprediction of Monarch. An example of such mispredition can be when Monarch predicts a block to be read only, but that is not the case and some writes happen to the block inside Monarch's scope. With this $t_{MWW}$ in place, Monarch can have $w_{WIN}$ writes to this single block without performance impact.}
Instead, we employ write mitigation techniques to alleviate the diverse impact of superset blocking.

\paragraph{Mitigating Writes}
Monarch incurs two kinds of writes into XAM arrays: writebacks from L3 due to eviction and block installation.
Most these writes have been proven unnecessary and counterproductive for in-package caches, where skipping them results in a better system performance~\cite{7284066,Accord}.
Similarly, Monarch eliminates a significant portion of these unnecessary writes to enhance the system lifetime.
%of which may be unnecessary with little/no performance increment, 
%causes faster wear-out of Monarch cells. To alleviate this issue, we write into Monarch only in specific conditions.}
%
%\textcolor{red}{
When fetching a missing block from memory, Monarch follows a \textit{no-allocate} policy, wherein no XAM memory will be allocated for the new block.
Hence, no writes is performed for a missing block.
% installations. 
Instead, the block is sent for installation in L3.
% directly sends the received block from the lower level to L3. }
This, however, may result in zero block installation in Monarch.
To address this issue, we allow a selective block installation in Monarch for the evicted blocks (dirty and clean) from L3.
%
%\textcolor{red}{To keep track of the nature of writebacks happening from L3, 
Along these lines, we add a new bit-flag per L3 blocks that indicates if the block is read after installation in L3.
%When a block is evicted, Monarch received this information plus the dirty bit information, along with the data, to help Monarch make an informed choice about writing the evicted block into it's arrays. If there is a high probability that the evicted block might not be used again in the near future, Monarch directly writes the block into main memory, thereby preventing unnecessary writes into it's arrays}
On a block eviction from L3, the dirty (D) and read (R) flags help the Monarch controller to install or update a block in the XAM arrays,
%\textcolor{red}{The nature of the dirty bit (D) and the read bit (R) results in 4 different scenarios of writeback into Monarch}
for which four different cases are possible.
$D\&R$:
If both bits are high, the block has been read from and written into by the program, which is likely evicted from L3 due to a capacity miss.
%Given that there is good activity for the block. 
We observed that it would help to write these blocks in Monarch.
%because the probability of this block being accessed again by the memory is high.
%
$D\&\overline{R}$:
This case indicates that the block has been written into, but not read from before eviction.
%there is a good chance that such blocks are just used in final value updates, meaning they might be \textit{dead} i.e never used again in the scope of the program. 
Monarch forward these blocks to the main memory. %with no Monarch write.
$\overline{D}\&R$:
In this case, the block has been read from, but not written into before eviction.
We observe most such blocks are read-frequently; therefore, Monarch identifies them as read-only and install them in XAM arrays.
%, meaning it is just a one-time write into Monarch, assuming that all other accesses to this block are reads.
%
$\overline{D} \& \overline{R}$:
If both bits are low, the block has not been accessed ever during its lifetime in L3.
These blocks are skipped from Monarch.
% after the initial installation
%For case (4), there is a good chance that such a writeback happens due to cache thrashing, where-in the working set size is greater than the size of L3. This usually occurs when a for loop iterates over a dataset of size greater than L3, in which every first access to a cacheblock within an iteration will be a miss in L3. It would help in this case to writeback the data into Monarch upon eviction given we avoid the added latency of accessing the main memory upon a miss
Overall, we observed an average of 31\% reduction in the write traffic to the in-package memory due to these rules.
%, these rule result in reducing the in-package write traffic by We observe about 30\% of the blocks are of this type, which are skipped by Monarch.

\paragraph{Distributing Writes}
%Previous work on improving lifetime \cite{QuEndu, Googlewear, changwear, 2003wear} has proposed wear leveling techniques that efficiently distributes writes across cells. Similiarly, Monarch adopts a low cost, yet effective wear leveling mechanism to increase the lifetime of XAM arrays.
%Our goal is to prevent fast wear-out in \textit{hot} parts of Monarch that are frequently written due to an uneven write traffic (or a malicious program).
%However, achieving this goal may require significant memory and performance overheads.
%To keep the costs low, 
% for distributing XAM writes within and across sets, supersets, banks, and vaults. %that achieves a nearly optimal lifetime for Monarch.
%For example, a
A cache placement/replacement policy may often end up choosing a few physical locations of a superset for eviction and installation.
To ensure an even distribution of writes within supersets and a fair use of the Monarch address space, we propose a rotary offset address mapping combined with a hierarchical mechanism.
%To avoid such write imbalance, w
We adopt a random replacement policy based on a free running 9-bit counter shared by all the sets within each vault.
On every block replacement, the vault controller increments the counter to indicate the next physical location of all sets for a possible eviction.
As a result, every two block evictions/installations at a physical location are spaced in time by at least 512 evictions per vault.

\begin{figure}[h!]
	\begin{center}
		\epsfig{file=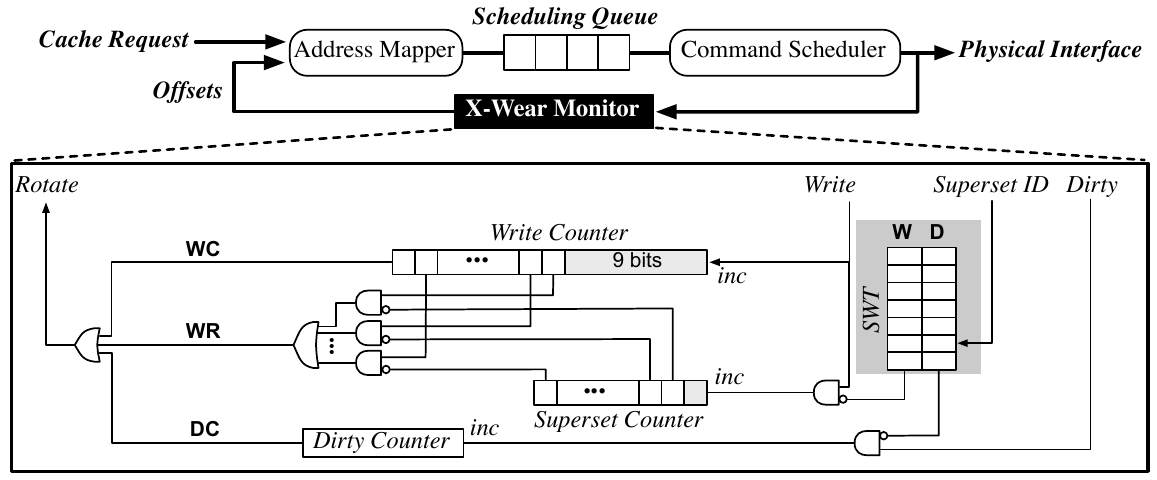, width = \columnwidth}
	\end{center}
	\caption{The proposed wear leveling logic at the cache controller.
		\label{figure:xrotor}}
\end{figure}

Moreover, Monarch employs a wear leveling mechanism at every vault controller (Figure~\ref{figure:xrotor}).
Similar to DRAM controllers, an address mapper and a command scheduler are used to (1) extract the bank, superset, row, and column IDs for each access, (2) track the required interfacing commands for each request while performing the necessary cache operations, and (3) enforce the required timing parameters among all the issued commands.
The proposed mechanism actively monitors all the tag updates, block installs, and data writes to each superset.
Upon detecting an uneven write distribution within each vault, a \textit{rotate} signal is generated to redistribute writes in Monarch through an offset address mapping.

We consider the average number of writes per superset (WR) as a metric to identify an uneven write distribution.
A large WR indicates that the accessed supersets are likely hot and candidates for remapping.
We employ a write counter and a superset counter for approximating WR.
Upon issuing every XAM write by the command scheduler, the write counter is incremented and a superset write table (SWT) is looked up for the W and D flags, which indicate if the superset has been written and dirty, respectively.
If this is the first write to the superset (i.e., W=0), the superset counter is incremented by 1 and W is set to 1.
If D is 0 and the write creates a dirty block in the superset, the dirty counter is incremented by 1 and D is set to 1.
To eliminate the need for a full-fledged divider, we employ a simple logic for approximating the ratio between the write and superset counts.
WR is set to 1 when the most significant non-zero bit of the write counter is 9 binary orders ($512\times$) larger than the superset counter.
The rotate signal is generated by ORing WR, WC, and DC; where, WC and DC indicate that the write and dirty limits have reached in the counters.
On a rotate signal, the dirty supersets from SWT are looked up and their dirty blocks are moved from Monarch to main memory, SWT and all counters are reset, and four address offsets are updated.
The offset registers are used for offsetting the vault, bank, superset, and set IDs of Monarch accesses during address mapping.
Upon a rotation, the offset registers are incremented by unique prime numbers--i.e., bank (1), set (3), vault (5), and superset (7)--to make the write distribution more even and less predictable.
(The vault offset is only updated every 8 vault rotates.)

%\subsubsection{Flat-RAM Mode}
%The proposed wear leveling hardware implement the start-gap~\cite{QuEndu} that requires two registers for counting the number of writes and keeping track of the number of gap movements.
%We use the superset counter as a gap register, thereby performing start-gap wear leveling with just an addition of a single gate for incrementing the superset counter for every time write counter crosses a threshold.
%Also, we employ the offset registers for randomizing the addresses as is proposed by start-gap~\cite{QuEndu}.

\section{Experimental Setup}
\label{section:setup}
%This section describes the system configuration, workloads, and methodology used for evaluating the Monarch and baseline systems.

\subsection{Architecture}
%For all of the evaluations, w
We consider an eight-core out-of-order CPU with two hardware threads per core using three levels of on-die cache.
L1 and L2 are private per core and L3 is shared.
Based on the ESESC multicore simulator~\cite{esesc}, we develop a heavily modified qemu-based cycle-accurate simulation platform that includes an accurate model of a DDR4 off-chip DRAM.%~\cite{ddr4}.
For the high bandwidth memory, we consider 8 vaults and the same number of TSVs for the SRAM, DRAM, RRAM, and XAM technologies.
We follow the WideIO protocol~\cite{jedechbm} for interfacing the memory stacks and CPU, but the timing parameters and command semantics are reevaluated based on each technology.
%However, the timing parameters are reevaluated for each technology, and an eight-vault Monarch interface.
We modify qemu to support a memkind extended library for memory allocation in the flat and cache modes.
%We implement three in-package memory models based on the DRAM, RRAM, and the new resistive XAM technologies.
%For the DRAM baseline, we consider four DDR4 DRAM layers~\cite{ddr4}.
%
%As mentioned before, Monarch vaults may be configured to operate in the hardware- or software-managed modes.
%A hardware-managed vault enables policies for automatic cache operations; while, a software-managed vault is a flat CAM or RAM memory.
For caching, we model three baselines: a DRAM-based set-associative cache~\cite{lohefficiently}, an iso-area SRAM cache and %as a baseline for comparison with the proposed XAM-based caching.
%To better represent the potentials of DRAM based caches, we also model 
an ideal DRAM cache with zero refresh, precharge, and activate overheads.
%Notice that Monarch can dedicate a complete superset for storing 512 (=$8\times64$) 64-bit keys, which makes the hardware-managed cache highly set associative.
%For example, caching 32-bit tags in a superset enables Monarch to realize a 1024-way set associative cache.
%(Notice that during the first step of a search transaction (row \textit{write}), a desired-length tag is repeated withing a 512-bit burst.)
%

To model the resistive in-package memories, we modify CACTI 7.0~\cite{balasubramonian2017cacti} by adding the necessary RRAM read/write components from NVSIM~\cite{li2016nvsim}. %\hl{Regarding XAM (2R) supply voltage, we understand that the numbers reported by differential 2R crosspoint ~\cite{chiu2014differential} will be an under-estimation of the required parameters, given that differential 2R crosspoint considers a local wordline of just 4 cells wide. 
We inspire numbers from the existing RRAM arrays~\cite{milo2019multilevel}: the read voltage of 1.0V and a write voltage of 2.2V.%bricalli2016siox,VoltageRRAM
%,wei2008highly
We assume that the write voltage is constant for every write across both resistors, overcompensation for the cases where no writes after sensing are required.
We further modify the environment to implement XAM arrays and control units for toggling the sensing reference and datapath at the superset and bank levels.
%We use the RRAM parameters from prior work~\cite{wei2008highly} to model the resistive elements in the arrays.
The system energy and power computation is done in coordination with McPAT~\cite{li2009mcpat} tool for the processor die, Micron power calculator~\cite{micron-ddr} for the main memory, and prior work on HBM memories~\cite{o2017fine} for the in-package cache architecture.
%Following the work on HBM memories~\cite{o2017fine}, 
We develop the necessary SPICE models for the set/superset/bank interconnects, multiplexers, and buffers in the modified CACTI 7.0 for accurate estimation of delay and energy.
All the circuit models for the newly added components are evaluated at the 22nm CMOS technology node~\cite{ptm}.
The results are then used for computing the interfacing parameters.
%In coordination with the developed simulation tool, we use an updated version of McPAT~\cite{li2009mcpat} for estimating the system power/energy.
Table~\ref{table:system} shows the system configurations.
\begin{table}[h!]
	\vspace{-4ex}
	\caption{System configurations.\label{table:system}}
	\vspace{-2ex}
	\centerline{
		\scalebox{0.7}{
			{%\scriptsize
				\begin{tabular}{|c|c|}
					\hline
					\multicolumn{2}{|c|}{\textbf{Processor}}\\
					\hline
					\textbf{Core} &  {8 OoO cores, 2 HW threads/core, 256 ROB entries, 3.2 GHz }\\
					\hline
					\textbf{IL1/DL1 cache} &  {64KB/64KB, 2-way/4 way, LRU, 64B block }\\
					\hline
					\textbf{L2 cache} &  {128KB, 8-way, LRU, 64B block }\\
					\hline
					\textbf{L3 cache} &   {8MB, 16-way, LRU, 64B block }\\
					\hline
					\hline
					\multicolumn{2}{|c|}{\textbf{In-Package DRAM}}\\
					\hline
					\textbf{Specifications} &  4GB, 8 layers, 8 vaults, 4 rank/vault,\\
					\textbf{}& 8 banks/vault, 1600MHz Wide I/O 2, 64 bits/vault\\        
					\hline
					\textbf{Timing} & {tRCD:44, tCAS:44, tCCD:16, tWTR:31, tWR:4, tRTP:46, tBL:4 }\\ 
					\textbf{(CPU cycles)} & {tCWD:61, tRP:44, tRRD:16, tRAS:112, tRC:271, tFAW:181 }\\
					\hline
					\multicolumn{2}{|c|}{\textbf{In-Package RRAM/Monarch}}\\
					\hline
					\textbf{Specifications} &  8GB, 8 vaults, 64 banks/vault, 256 supersets/bank, 8 sets/superset 64 rows/set\\
					\textbf{}& 32 banks/vault, 1600MHz Wide I/O 2, 64 bits/vault, $R_{lo} = 300K$, $R_{hi} = 1G$\\        
					\hline
					\textbf{Timing} & {tRCD:4, tCAS:4, tCCD:1, tWTR:31, tWR:162, tRTP:1, tBL:4 }\\ 
					\textbf{(CPU cycles)} & {tCWD:4, tRP:8, tRRD:1, tRAS:4, tRC:12, tFAW:181 }\\
					\hline
					\multicolumn{2}{|c|}{\textbf{In-Package CMOS (CAM+RAM)}}\\
					\hline
					\textbf{Specifications} &  73.28MB, 8 layers, 8 vaults (ISO-area to 8GB Monarch)\\        
					\hline
					\textbf{Timing} & {tRCD:4, tCAS:4, tCCD:1, tWTR:31, tWR:3, tRTP:1, tBL:4 }\\ 
					\textbf{(CPU cycles)} & {tCWD:4, tRP:8, tRRD:1, tRAS:4, tRC:12, tFAW:181 }\\
					\hline
					\multicolumn{2}{|c|}{\textbf{Off-Chip Main Memory}}\\
					\hline
					\textbf{Specifications} &  32GB,  2 channels, 1 ranks/vault, \\
					\textbf{}& 8 banks/rank,1600 MHz DDR4, 64 bits/vault\\
					\hline
					\textbf{Timing} & {tRCD:44, tCAS:44, tCCD:16, tWTR:31, tWR:4, tRTP:46, tBL:10}\\ 
					\textbf{(CPU cycles)} & {tCWD:61, tRP:44, tRRD:16, tRAS:112, tRC:271, tFAW:181 }\\
					\hline
			\end{tabular}}
		}
	}
	\vspace{-3ex}
\end{table}

\subsection{Workloads}
%We consider in-package caching, hashing, and searching to assess the performance potentials of Monarch for the big data processing applications.

\subsubsection{In-Package Hardware-Managed Caching}
%We assess the power and performance potentials of the proposed and baseline systems by executing 
We simulate a mix of 11 parallel applications from the CRONO benchmark suite~\cite{ahmad2015crono} for scalable graph processing and memory intensive NAS Parallel Benchmarks~\cite{NASbench}.
%, SPLASH-2~\cite{woo.isca95} and Phoenix~\cite{yoo.iiswc09} benchmark suites.
%The serial applications are selected from integer and floating point SPEC2017 benchmark suites~\cite{spec2017}.
%CRONO: A Benchmark Suite for Multithreaded Graph Algorithms Executing on Futuristic Multicores
%All of the selected serial applications are executed in the rate mode.
We use \texttt{GCC} to compile all of the applications with \texttt{-O3} flag.
%For the serial benchmarks, we consider warming up the cache by fast forwarding the initial instructions until the cache is full; then, we simulate the next one billion instructions or until the application completes, whichever happens sooner.
%For the parallel applications, we execute every program two times to completion, one run immediately after another.
%The first run starts with an empty cache.
All of the programs are executed to completion.
%For the second run, we ensure that all the virtual to physical mappings are different from those in the first run.
%We consider the results of the second run for each application.
%Table~\ref{table:applications} shows the workload characteristics and their corresponding input sets.
We choose inputs of class A for all the evaluated NAS benchmarks, namely Fourier Transform (FT), Conjugate Gradient (CG), and Embarrassingly Parallel (EP).
The CRONO benchmark applications, however, allow the user to determine the input size.
We configured all of the graph applications to process input graphs that generate a footprint at least $2\times$ the size of the in-package memory.
The evaluated CRONO benchmark applications are Betweenness Centrality (BC), Breadth First Search (BFS), Community Detection (COM), Connected Components (CON), Depth First Search (DFS), Page Rank (PR), Single Source Shortest Path (SSSP), and Triangle Counting (TRI).
%We will extract execution time, DRAM cache and system energy, hit rate, and aggregated traffic on both HBM cache main memory from each run.
%In addition, parallel benchmarks runs to completion two times.  
%
%
%\begin{table}[h!]
%	\vspace{-3ex}
%	\caption{Workloads and data sets.\label{table:applications}}
%	\vspace{-2ex}
%	\centerline{
%		\scalebox{1}{
%			{\scriptsize\setlength\tabcolsep{3pt}
%				\begin{tabular}{|c|c|c|c|}
%					\hline
%					\textbf{Label}  &\textbf{Benchmarks}  & \textbf{Suite} &\textbf{Input} \\
%					\hline
%					\hline
%					FT &Fourier Transform & NAS  & Class A \\ 
%					%MG &Multi-Grid & NAS & Class A \\
%					CG & Conjugate Gradient
% & NAS & Class A \\
%					EP & Embarrassingly Parallel
% & NAS & Class A \\
%					%LU & Lower-Upper Gauss-Seidel solver & NAS & Class A \\
%					\hline
%					BC
%&  Betweenness Centrality & CRONO & 2$\times$ \\
%					BFS
%& Breadth First Search & CRONO & 2$\times$ \\
%					COM
%& Community Detection & CRONO & 2$\times$ \\
%					CON
%& Connected Components & CRONO & 2$\times$ \\
%					DFS
%& Depth First Search  & CRONO & 2$\times$ \\
%					PR
%& PageRank & CRONO & 2$\times$ \\
%					SSSP
%& Single Source Shortest Path & CRONO & 2$\times$ \\
%					TRI
%& Triangle Counting & CRONO & 2$\times$ \\
%					\hline
%			\end{tabular}}
%		}
%	}
%	\vspace{-2ex}
%\end{table}

\subsubsection{In-Package Software-Managed Hashing}
We implement Hopscotch Hashing  with a Murmur3  hash function as the main kernel of analysis for Monarch. Hopscotch Hashing is an open addressing method of hashing, whose performance is scalable with the hash table size. This scalability is achieved through usage of windowing, which reduces the search space for a specific key from the whole hash table to just a window of buckets.
%\cite{murmur}, \cite{hopscotch}
Hash table lookup finds the intended index of the key through a hash function. It then uses metadata to identify which of the buckets in windows \textit{i} to \textit{i + window\_size} have keys indexed to the same index.
Then, the algorithm checks for a match among them, which may be accelerated through Monarch search.
Monarch combines multiple iterative read requests into a single search request.
The insert process performs a lookup to see if the key is already present. If not, it iterates through the size of the hash table window unless it finds the next free bucket. If this free bucket is present within the window size, we store the new key there. If not, it iterates through the rest of the hash table to find an empty bucket. If it finds one, it tries to do some swapping of keys such that the rule of the hash table is preserved. If the rule is not satisfied, it keeps rehashing the table to a higher size and retries the insert till it succeeds.

%We consider 4 baselines for the evaluation, DRAM HBM as the L4 cache, RRAM used for the in-package memory with the updated timing parameters of WideIO,  CMOS-CAM/RAM as explained in section~\ref{section:technology}DRAM HBM configured for operation in the hybrid mode, which allows the in-package memory to be used partially for software defined scratchpad memory and hardware-managed cache
%, similiar to the Intel's KNL. %This offers a fair comparison owing to the similiar bandwidth of operation to Monarch.
To preserve the order of search requests generated by the application, all of the memory accesses in the Monarch CAM address space are set to non-cacheable for on-die memories.
%\textcolor{red}{Unlike KNL, Monarch's user defined address spaces don't allow on-chip caching owing to coherence required for doing the search operations}.
%
%Table~\ref{table:paraX} shows the simulation parameters considered for the baselines and Monarch architectures.
%
%For baseline 1, the processor is connected to DRAM memory interfaced with HBM, which is connected to DDRx. The HBM is configured to be in Cache mode 
%
%For baseline 2, the processor is connected to RRAM memory interfaced with WideIO. which is connected to DDRx.
%
%For baseline 3, the processor is connected to DRAM memory interfaced with HBM, which is connected to DDRx. The HBM is configured to be in Hybrid mode, with allows for Software defined data placement. 
%
In both RRAM and KNL-like HBM, software controls data placement using \textit{HBM\_malloc} and \textit{flat-RAM\_malloc}. Data not belonging to this address space are allocated in DDRx. All memory requests excluding the ones specified by the aforementioned mallocs are cacheable.   
For the SRAM system, if the working set size exceeds the storage limit, the remaining part of the working set is stored in main memory.

We benchmark our tests for hashing predominantly on the YCSB-B \cite{YCSB} microbenchmark, which specifies a zipfian-distributed workload with a 95\%/5\% read/write distribution respectively to the hash table. Sensitivity analysis is done across density, read/write percentage, window size, and Hash Table size to further evaluate the performance of Monarch across all possible characteristics of the hash table.

\subsubsection{In-Package Software-Managed Searching}
An important class of big data processing relies on large scale search.
For example, we evaluate in-package search for string matching on the proposed Monarch and baseline systems.
We choose the parallel version of String-Match kernel from the Phoenix \cite{yoo2009phoenix} for the evaluations.
The String-Match kernel is used in various applications such as gene sequence analysis \cite{liu2005fast}.
%This evaluation provides insight into the size v/s performance tradeoff that Monarch offers, where-in the strings have to be block aligned, which enables users to use the search functionality effectively
%

\section{Evaluations}

%\subsection{Methodology}

%We use the ESESC~\cite{esesc} simulator for modeling a multicore processor system with three on-die cache levels. 
%To compute the per-access energy overhead of the proposed Monarch controller, we use CACTI 7.0~\cite{balasubramonian2017cacti}.
%The system energy and power computation is done using the ESESC simulator~\cite{esesc} in coordination with McPAT~\cite{li2009mcpat} tool for the processor die, Micron power calculator~\cite{micron-ddr} for the main memory, and prior work on HBM memories~\cite{o2017fine} for the in-package cache architecture.

\subsection{Hardware Overhead}
%To assess the potentials and overheads of the proposed wear leveling mechanism, we model all the additional parts in our simulation environment.
For the {8GB Monarch, SWT is 8KB.}
We also consider another 4KB for buffering the $t_{MWW}$ counts on-chip.
We use the SRAM and logic components from CACTI~\cite{balasubramonian2017cacti} to estimate the area, delay, and energy of the additional hardware.
Together, the SRAM buffers incur an insignificant overhead to the chip area of a KNL-like processor (less than 2\%).
As the additional control for wear monitoring is performed in parallel with command scheduling, none of the interfacing parameters are affected.
However, address remapping for every incoming request is delayed by one cycle, which is accurately considered in all of our evaluations.

\subsection{Monarch in Cache Mode}
\begin{figure}[h!]
	\begin{center}
		\epsfig{file=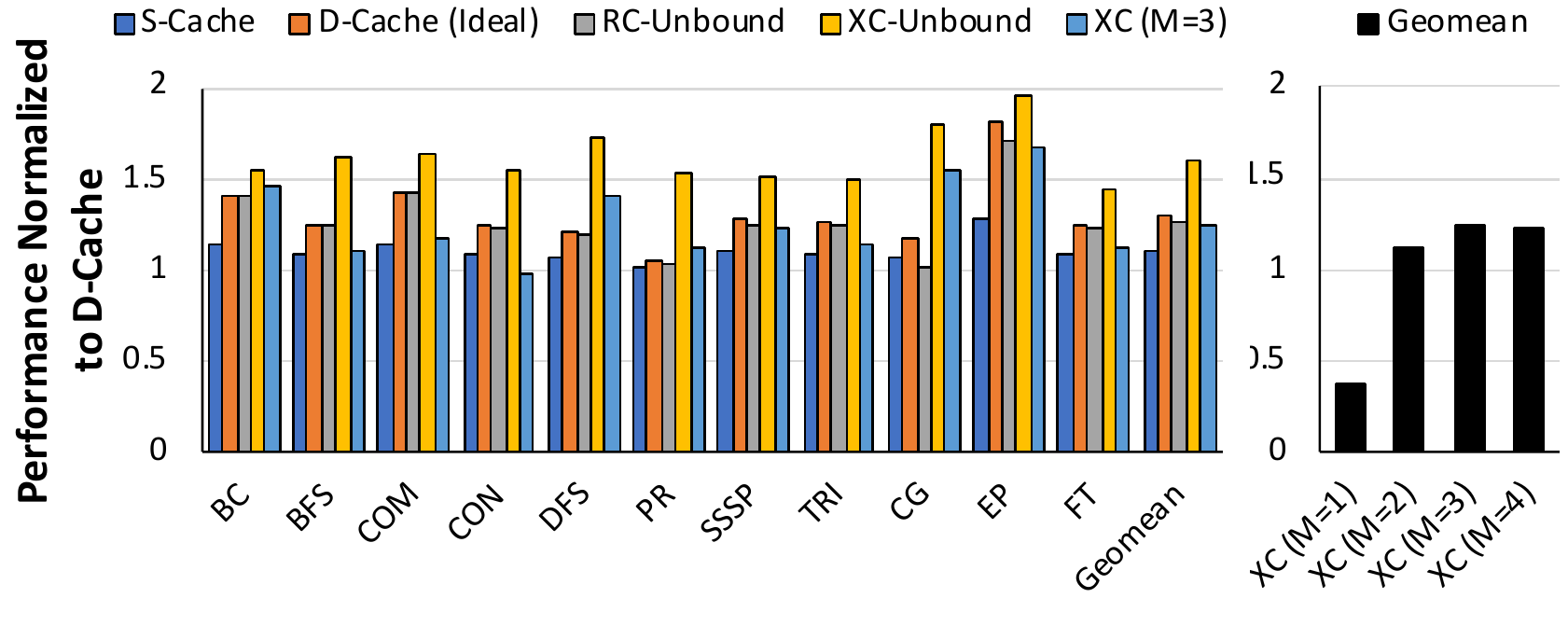, width = \columnwidth}
	\end{center}
	\caption{Performance of the Monarch and baseline systems.
		\label{figure:xcache_performance}}
	
\end{figure}

We consider an unbound version of Monarch without considering any constraints such as $tMWW$ and wear monitor, as well as multiple bounded version of Monarch with respect to the number of block writes permitted per window ($M=1,2,3,$ and $4$).
Also, we set the target lifetime ($T_{Life}$) to 10 years~\cite{server_lifespan}.
% as a minimum lifetime guarantee.}
We compare the results with in-package caches using SRAM, DRAM, Ideal DRAM, and unbound RRAM.
As shown in Figure~\ref{figure:xcache_performance}, we observe an average performance improvement of 61\% for the unbound Monarch compared to the DRAM cache, which is 21\% more than the attainable performance improvement for D-Cache (Ideal).

The unbound RRAM based cache (RC-Unbound) can achieve an average performance gain of 24\%, which is similar to that of the D-Cache (Ideal) due to eliminating the needs for activate, precharge, and refresh.
When $t_{MWW}$ is enforced, Monarch encounters a performance loss.
{The loss is more significant for $M=1$.
As we increase $M$, more frequent writes are permitted per write window.
This also increases the amount of time a superset may be blocked after exceeding the number of writes.
For the evaluated applications, $M=3$ provides a slightly better performance improvement (25\% on average) rather than the others.
S-Cache, the reconfigurable 3D stacked CMOS, provides a faster access but significantly less capacity compared to the counterpart technologies.
As a result, its performance is better than D-Cache but worse than the other evaluated systems.
}

We observe similar trends in energy savings due to reducing the static power and DRAM overheads.
Monarch (M=3) exhibits an average of 21\% reduction in system energy; while, RRAM reduces 13\% of the overall system energy.
We also observe that the higher associativity of Monarch results in a significant hit-rate improvement for some applications.
For example, Monarch (M=3) gains more than $2\times$ boost in hit-rate for BC.
(RC-Unbound and D-Cache implement the same cache architecture in different technologies; therefore, they exhibit similar hit-rates.)

It is interesting to note that performance improvement does not necessarily correlate with the hitrate for some applications such as BC and PR. This relies on the fact that TSVs offer higher bandwidth, but not necessarily a lower latency in comparison to the off chip interface.

A tag check requires fetching a tag from 3D layers to the cache controller. Depending on if there is a hit or a miss, a read request is either sent to the main memory or to the 3D layers, both of which may have a similiar read latency. Hence, if the application does not saturate the bandwidth of the main memory, there will virtually be no performance difference in the system irrespective of the hitrate.
\begin{figure}[h!]
	\begin{center}
		\epsfig{file=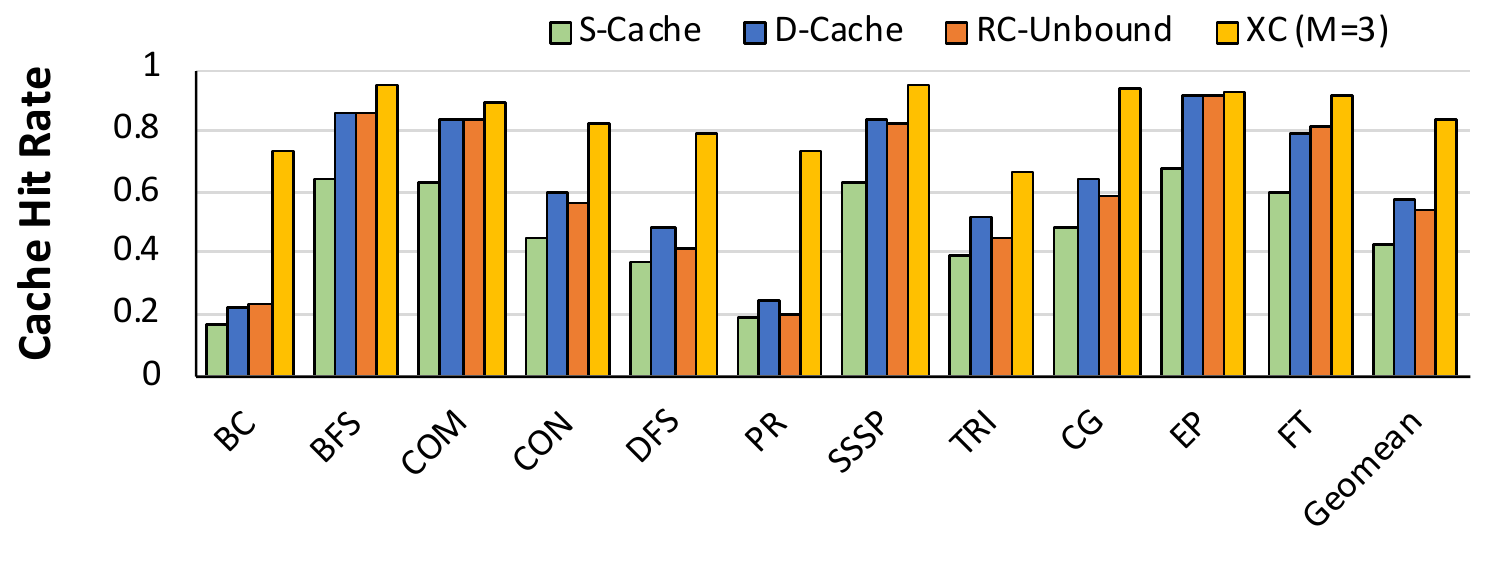, width = \columnwidth}
	\end{center}
	\caption{Cache hitrates for the RRAM and DRAM baselines vs. the proposed Monarch.
		\label{figure:xcache_hitrate}}
\end{figure}

\subsection{Lifetime Assessment}
In addition to performance evaluations, we performed specific simulations for lifetime assessment.
Here, we explain the summary of lifetime assessment for $M=3$.
For this experiment, we simulate all the multithreaded applications to the end of execution while recording Monarch snapshots at every rotation.
On average, rotates happen at every 260M cycles.
To limit the writeback overheads of dirty blocks on every rotate, we set DC to 8192 (Section~\ref{section:lifetime}).
The maximum performance drop due to dirty block writeback is less than 1\%.
However, we observe an average of less than 4\% performance degradation due to the increased cache misses after flushing Monarch at every rotation.

\begin{figure}[h!]
	\begin{center}
		\epsfig{file=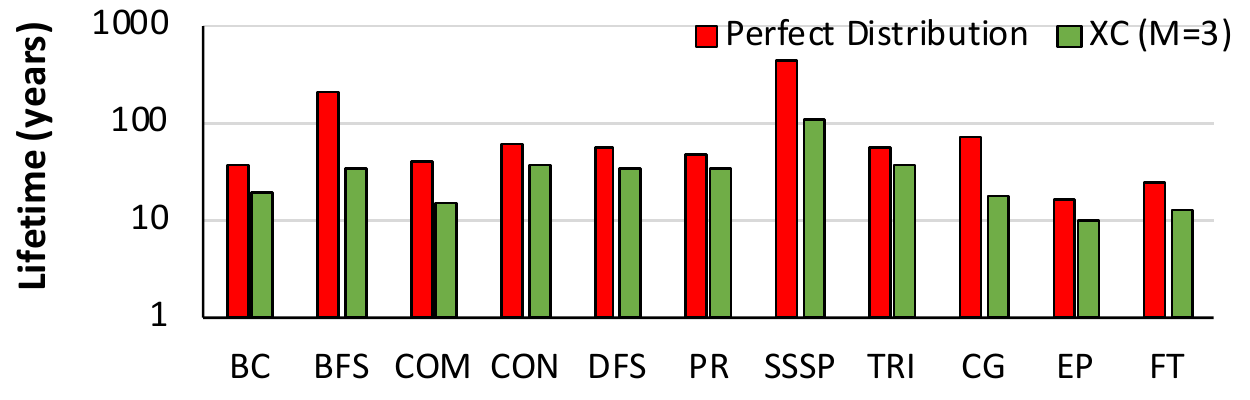, width = \columnwidth}
	\end{center}
	\caption{Enhanced lifetime for Monarch.
		\label{figure:lifetime}}
	\vspace{-3ex}
\end{figure}

Performing a cycle accurate simulation till RRAM cells die seems impractical for estimating lifetime.
Instead, we use the recorded memory snapshots for lifetime estimation.
Each recorded snapshot contains the number of writes performed to each XAM row and column.
We model a constantly repeated execution of each application while applying the offset addressing on every rotation.
The lifetime estimation stops when a XAM cell exceeds the maximum number of cell writes (e.g., $10^8$).
Figure~\ref{figure:lifetime} shows the lifetime of Monarch ($M=3$) using the proposed wear leveling mechanism compared to an ideal wear leveling.
The ideal lifetime is theoretically computed based on the produced write bandwidth by each application and a perfectly even distribution of writes in Monarch.
The minimum lifetimes are observed for EP with 16.72 and 10.22 years for the ideal and Monarch ($M=3$) systems. %, respectively.

\begin{figure*}[h!]
	\begin{center}
		\epsfig{file=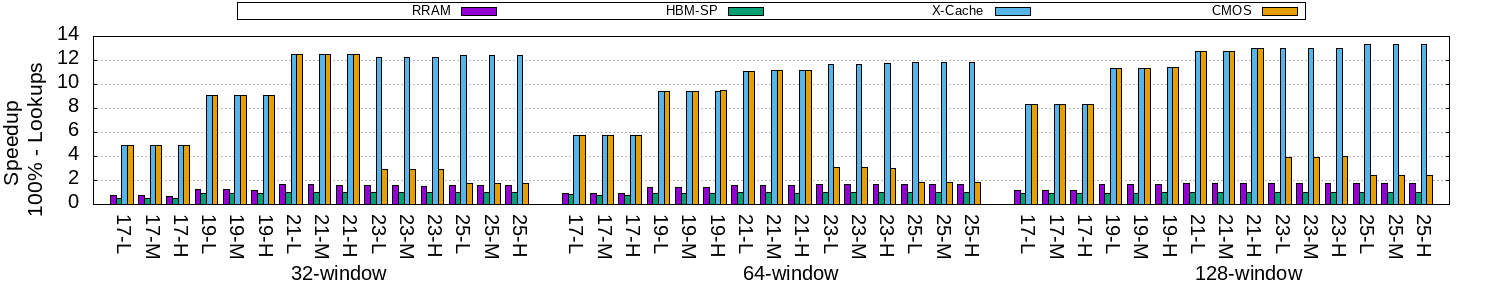, width = 2.2\columnwidth}
	\end{center}
	\caption{Performance of all configurations relative to HBM-C for 100\%-Reads
		\label{figure:100L}}
\end{figure*}

\begin{figure*}[h!]
	\begin{center}
		\epsfig{file=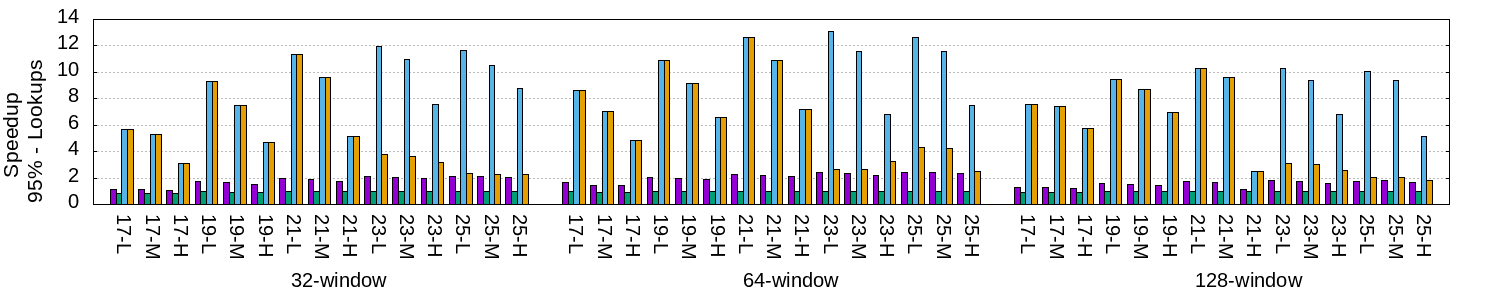, width = 2.2\columnwidth}
	\end{center}
	\caption{Performance of all configurations relative to HBM-C for 95\%-Reads
		\label{figure:95L}}
\end{figure*}

\begin{figure*}[h!]
	\begin{center}
		\epsfig{file=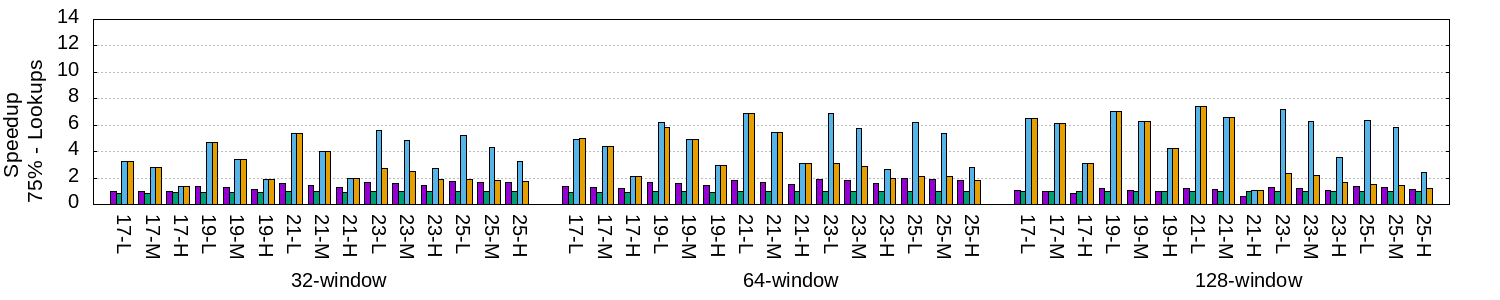, width = 2.2\columnwidth}
	\end{center}
	\caption{Performance of all configurations relative to HBM-C for 75\%-Reads
		\label{figure:75L}}
\end{figure*}

\subsection{Monarch in Flat Mode for Hashing}

\subsubsection{Lifetime Consideration}
%\textcolor{red}{
Monarch in flat mode uses $t_{MWW}$ to ensure a minimum lifetime guarantee for Monarch, which may result in significant performance degradation for write-intensive data. Fortunately, Hash Table accesses are inherently read intensive, so that keeps the write bandwidth low. Rehashing is naturally done within the scope of main memory to prevent performance drop incurred by $t_{MWW}$. On rehashing, the new hash table is copied into Monarch. This overhead is considered as part of our evaluations
%} 

\subsubsection{Impact of Window Size}
We study the maximum density of hash tables before resizing for different window sizes.
We observe that with increasing the size of a hash table, the difference in the maximum number of keys could be stored in the table across different window sizes increases.
%
%Based on our observation regarding the maximum density of the hash table before resizing for different window sizes, the difference between the maximum number of keys a hash table of a specific size could hold before resizing for increasing window sizes increases with the size of the hash table.
%Our analysis on the hash tables indicate that increasing the size of hash table ...
%difference between maximum number of keys before resizing across different hash table window sizes increases with increasing size of the hash tables
%
Moreover, the probability of rehashing upon insertion decreases as the window size increases.
This is a desirable quantity as resizing is basically re-insertion of all keys from the old hash table to the new hash table of a higher size.
(Insertions are costly and high-latency operations.)
However, increasing the window size results in storing more metadata per bucket ($\textit{window\_size}/8$).
While the metadata is used for lookups in the baseline, Monarch does an index based search on the XAM arrays, effectively deeming metadata unnecessary for lookups. Hence, metadata can be stored in the main memory , thereby freeing up much needed space in Monarch for more key-value/key-pointer pairs.
The baselines considered for comparison are RRAM (Monarch as pure flat-RAM), HBM-C (DRAM HBM as L4 in pure cache mode), HBM-SP (DRAM HBM in pure scratchpad mode) and CMOS (SRAM based model as described in section~\ref{section:technology})
 
\subsubsection{Hashing for 100\% Lookups}
Figures~\ref{figure:100L} shows the {performance of Monarch, CMOS, RRAM, and HBM-SP over HBM-C for 32, 64 and 128 window sizes respectively.} The window size has minimal impact on performance owing to the nature of lookups, with each key being found within the first few buckets of a hash. The stagnation of Monarch relative performance at higher working set sizes is due to the diminishing impact of caching on performance of the baselines as size increases.

\subsubsection{Hashing for 95\% Lookups}
For 95\% Reads (YCSB-B), relative performance of Monarch increases with increasing the window size due to the complexity of lookups and inserts. The complexity of inserts increases with density, which leads to the diminishing relative performance of Monarch over the baselines. A similar trend to the 100\% lookup case is observed with the impact of increasing working set size

\subsubsection{Hashing for 75\% Lookups}
For 75\% reads, a similar trend as that of 95\% reads follows. One major difference is the relative performance wrt RRAM that comes closer to the relative performance wrt HBM-C/HBM-SP, owing to the higher percentages of writes.

Even though CMOS has a $50\times$ better write latency than Monarch, they perform similarly when the working set fits inside the CMOS stack.
This is attributed to (1) the low frequency of writes in general, (2) the write latency can be hidden if the bank to which write is happening has no pending critical request, which is very common for the random access pattern of hashing applications, and (3) the TSV interface bottlenecks the write bandwidth, leading to diminishing effects of the SRAM technological advantage.
For 23 and 25 window sizes, the respective size of key + data is 128MB and 512MB, which are bigger than the SRAM capacity.
This overflow of data results in a steep degradation of performance, in comparison to the 17, 19, and 21 window sizes.

\subsubsection{Best/Worst Case Performance of Monarch}
The relative performance of Monarch goes down heavily with increasing the percentage of key inserts. Higher key inserts lead to lower lifetime of Monarch owing to much more frequent writes happening into the memristor cells. But, many hash table applications follow the general trend of having a very low frequency of inserts. Our analysis on Wordcount~\cite{yoo2009phoenix} shows that the ratio of lookups to inserts would be 94:6, if a hashmap were to be used for MapReduce.
{Moreover, for general hashing applications such as Facebook's Memcached, analysis shows that the GET/SET ratio is 30:1 (3.33\% writes)}{~\cite{memcachedanal}}.
The general case performance is established in the aforementioned analysis, with every lookup being a hit and every insert being a new key. Offline analysis for the outlier case where a hash table lookup happens for keys that isn't present in high window size hash table presents the best case improvement of Monarch. Such an analysis with 100\% lookup on a 128-Window hash table of size 25 showed that Monarch outperforms HBM-SP by $54\times$ and $70\times$ in a low density and high density setting respectively.
\iffalse
\begin{figure}[h!]
	\vspace{-2ex}
	\begin{center}
		\epsfig{file=plots/Energy_Improvement.pdf, width = \columnwidth}
	\end{center}
	\vspace{-2ex}
	\caption{Energy improvements in the stacked memory using Monarch.
		\label{figure:energy_improvement}}
	%\vspace{-6ex}
\end{figure}
\fi

\subsubsection{Energy of Monarch}
The energy improvement in the 3D stacked memory is measured for the 75\% lookup case, owing to that having the most number of writes and hence, worst relative energy to HBM-SP. With increasing density, the energy consumption increases owing to the increasing number of writes. The trend with respect to increasing the window size depends on the nature of keys being looked-up and more importantly, inserted into the hash table.
Our analysis shows that Monarch achieves energy improvements between $2.4\times$ and $2.8\times$ over HBM-SP for the 75\% lookup case.

\subsection{Monarch in Flat Mode for Searching}
\iffalse
%\subsubsection{String Match} 
\begin{figure}[h!]
	%\vspace{-2ex}
	\begin{center}
		\epsfig{file=plots/String_Match.pdf, width = 0.9\columnwidth}
	\end{center}
	\vspace{-2ex}
	\caption{Performance of Monarch relative to baselines for StringMatch
		\label{figure:SM}}
	%\vspace{-2ex}
\end{figure}
\fi
String-Match~\cite{yoo2009phoenix} requires a large scan over the dataset for some target strings. This enables Monarch to broadcast large scale searches across the span of the dataset, with each search covering upto 4KB of data. This drastically reduces the number of requests that have to be sent to the memory.
We consider data storage from DDRx into Monarch, where data must be copied from DDRx into the CAM arrays, with each 64-bit CAM block boundary acting as the word delimiter.  This adds a two-fold overhead to the data storage in Monarch, wherein pre-processing is done to block-align the data, which in-turn results in an $8\times$ increase in data size. With the above overheads in consideration, Monarch performs $14\times$, $12\times$, $11\times$ and $24\times$ better in String-Match over RRAM, HBM-C, CMOS and HBM-SP respectively.
{The working set size in consideration is 500MB, which explains the relative performance of CMOS being similar to RRAM.}

% 
\iffalse
\begin{figure}[h]
	\begin{center}
		\epsfig{file=plots/resize.pdf, width = \columnwidth}
	\end{center}
	\caption{No of Inserts v/s Rehashing
		\label{figure:resize}}
\end{figure}
\fi

\iffalse
RRAM cells have limited write endurance. This usually translates to a low lifetime of memory. Previous work on improving lifetime \cite{QuEndu, Googlewear, changwear, 2003wear} has proposed wear leveling techniques, which incorporates microarchitectural techniquies for efficient write distribution across cells. Though resistive memories are under development and high endurable cells are being exemplified such as \cite{luo2019vlsit}, we incorporate a few software features and constraints to improve endurance.  
\fi

\iffalse
For caching, due to the high associativity of the cache, block replacement due to conflict misses has become less likely.
However, tag/data updates are inevitable.
We incorporated a free running counter at the controller that increments by 1 per every block update.
The counter is used as a pointer to replace/place the cache blocks in set.
We noticed this simple mechanism reduced the worse case life of of the sets by a factor 6 across all of the evaluated applications.
\fi

\section{Related Work}

%\subsection{Background}

\paragraph{High Bandwidth DRAM}
HBM~\cite{jedechbm} is in-package 3D stacked DRAM that has 8 DRAM dies stacked on top of each other. Through silicon vias (TSVs) connect the DRAM layers through the cross section, and the memory follows a hierarchy of vaults, banks, sub-banks, and sub-arrays. Each HBM typically has 4-8 vaults% similar to Figure~\ref{figure:layout}
, with each vault having 8-16 banks. %The high number of HBM banks (up to 512) offers high data parallelism, which can effectively utilize the high bandwidth provided by the HBM interface. 
The HBM interface allows for different DRAM commands such as \textit{refresh}, \textit{precharge}, \textit{activate}, \textit{read}, and \textit{write}.
%Each HBM vault has a dedicated memory controller that orchestrates data movement the processor and DRAM layers.   
%
%
%\paragraph{Intel's MCDRAM}
An in-package multi-channel DRAM (MCDRAM) is employed by 
Intel's Xeon Phi processors (code-named as Knights Landing)~\cite{mcdram}.
% to increase bandwidth available to an application. This DRAM is present on the same level in the hierarchy as the conventional DDR main memory. 
MCDRAM supports 3 operational modes for flat, cache, and hybrid address spaces.
The addressing mode configuration is performed at boot time.

%\begin{table} [h!]
%	\vspace{-2ex}
%	\caption{MCDRAM Modes in Knights Landing}
%	\vspace{-2ex}
%		\centerline{
%		\scalebox{1}{
%			{\scriptsize\setlength\tabcolsep{3pt}
%			\begin{tabular}{| c | l |}
%			\hline
%			Mode & Description \\ [0.5ex]
%			\hline
%			Cache mode & MCDRAM configured as L4 HBM cache to DDRx \\
%			\hline
%			Flat mode & MCDRAM configured as memory extension to DDRx \\
%			\hline
%			Hybrid mode & Partitioned as cache (25\% or 50\%) and  memory (rest) \\
%			\hline
%		\end{tabular}}
%		}
%	\vspace{-2ex}
%	}
%\end{table}

%\subsection{Related Work}

\paragraph{Reconfigurable Memory}
Software defined memories have been examined by the prior work.
%caches have been proposed previously by researchers. 
Jenga \cite{jenga} is a software defined cache that builds virtual cache hierarchies depending on the OS needs. Banshee \cite{banshee}  exploits software hardware interplay by combining page-table based page mapping management and bandwidth efficient cache replacement.
Li-Si Monona \cite{Zha2018} proposes a reconfigurable computing fabric coupled with RRAM based memory to alleviate FPGA performance and energy efficiency of diverse cloud workloads. 
Monarch provides configuration at a higher level, where the OS can dictate which part of the application needs to run with Monarch as cache, while it can configure a part of Monarch as software defined scratchpad and control data placement with the full address space of the scratchpad being visible.

\paragraph{Database on HBM}
Research on stream processing with HBM has largely focused on comparing sort merge versus hash join performance. StreamBox-HBM \cite{StreamBox} proposes a stream analytics engine optimized for HBM, wherein sequential sort algorithms for grouping are better than hash based join algorithms.
Monarch optimizes stream processing applications and operates at the same maximum bandwidth of HBM.  
Recent work on optimizing HBMs for hash tables have focused on alleviating the overhead of pointer chasing  \cite{pointerchase}, which uses the logic layer of the HBM to perform some in memory commutations for pointer chasing, thereby alleviating the data transfer cost from CPU to memory. This provides optimization for hash tables with separate chaining based conflict resolution. While Monarch implicitly reduces data transfers from memory to CPU for hash table lookups, it focuses on hash tables with open addressing based conflict resolution, which requires no pointer chasing. 
%
%
%\paragraph{High-Bandwidth Non-Volatile memory.}
R-Cache \cite{rcache} proposes a highly set-associative, in-package cache made by 3D die stacking of memristive memory arrays. However, R-Cache is purely hardware managed and has no reconfigurability, and the transaction flow differs between Monarch and R-Cache.
Moreover, R-Cache does not provide any lower bound for the lifetime; whereas, Monarch ensures a minimum lifetime regardless of the user applications.
Also, Monarch benefits from skipping unnecessary writes to improve performance.

\paragraph{Database in Non-Volatile Memory (NVM)}
Prior work on NVM for hash tables \cite{levelhash} has been in building novel single-node hash table structures to benefit from the non-volatility and low latency of NVM.  This work proposes a new memristive memory architecture, which can be exploited to optimize open addressing based single-node hash tables. HiKV \cite{hikv} focuses on designing a hash table that performs a hybrid indexing that exploits the fast index searching capability of NVM and low read latency of DRAM. 
NVMcached \cite{nvmcached} designs an NVM based key-value cache using byte-addressable NVM for key-value store. A maximum throughput of $4.8\times$ on YCSB workloads is gained while Monarch achieves a maximum speedup of around $13\times$ for key-value search.
%
%FaRM and HERD \cite{farm,herd} use hopscotch hashing to improve distributed systems hash table performance using RDMA. Monarch is focused on improving performance of single-node hash tables.
%
RC-NVM \cite{rcnvm} proposes a {1R crosspoint based memory architecture} that supports row and column accesses to memory allowing for optimizations of iMDB based applications like SQL. RC-NVM also proposes new instructions to enable row/column storage of data. {Monarch proposes a row/column storage similiar to RC-NVM, but unlike RC-NVM, provides the capability of large scale associative search, while also providing a 2R-cell based crosspoint which is more robust against possible sneak currents due to half selected cells \cite{xu2015overcoming}}.

\paragraph{Persistent Memories}
Intel Optane 3D XPoint features a crosspoint structure made of fast switching material. Optane DC \cite{optaneanalysis} supports memory (cached) and app direct (uncached) modes. In memory mode it is used as cache for DDRx, which helps hide the drawbacks of the low bandwidth of DDRx, while in App Direct mode, it is used as a persistent storage device, similiar to Monarch. Optane has been shown to be almost $2\times$ faster than their counterparts for file-system storages. However, Monarch is different due to its multidimensional access and search capability.

\paragraph{Lifetime Enhancement of NVM}
Wear levelling for different technologies have been proposed previously which applies across the range of non-volatile memory technologies. Qureshi ~\cite{QuEndu} et al. proposes start-gap wear leveling, which uses only a start register and a gap register coupled with simple address-space randomization with minimal overhead. To the best of our knowledge, Monarch is the first memory system providing a use-defined lower bound for resistive caches.
%Industry approaches have previously proposed wear leveling for flash EEPROM \cite{2003wear}. Chang ~\cite{changwear} proposes wear leveling for flash memories through a two-level dual-pool algorithm. Yeh et al. \cite{Googlewear} proposes a block-based wear leveling with spare blocks.

%\paragraph{Associative search based architectures.}
%Architectures that allow for associative search based on TCAMs have been proposed in the past for different applications \cite{agrawal1994fast, meribout2000using, guo2013ac, de2000needles, ranger2007evaluating}. Monarch is RRAM based and works at a much finer granularity than the previous works. The X-WideIO interface also allows for a much higher bandwidth than the previous works, an instance of which can be noted at the difference in performance of string match between Monarch and AC-dimm \cite{guo2013ac}.

\section{Conclusion}
{
This work presents Monarch, a memristive 3D stacked memory based on a novel array structure called XAM.
The XAM array provides low-cost capabilities for runtime configurations between content addressable and random memory modes.
As compared to DRAM, XAM eliminates the need for costly DRAM operations (such as activate, precharge, and refresh), enables a denser memory, improves tag management, and allows for in-package memory acceleration.
% in high-bandwidth memories and provides a denser  incurs much less overhead in comparison to HBM which helps us better utilize the bandwidth available to the memory. To effectively utilize the reconfigurability provided by Monarch, this work provides some OS primitives to enable user-defined address allocation in specific configurations of the memory. 
Monarch exhibits significant potential to improve the performance and energy-efficiency of data-intensive applications using associative searching, software-managed in-package buffering and hardware-managed caching.
% based application, while also allowing for usage as general purpose highly set-associative in-package cache.
}

\ifCLASSOPTIONcaptionsoff
  \newpage
\fi

\bibliographystyle{IEEEtran}

\bibliography{references}

\end{document}